\documentclass[sigplan,10pt,preprint]{acmart}

\usepackage[english]{babel}
\usepackage{blindtext}
\usepackage{subfig}
\usepackage{xcolor}
\usepackage{colortbl}
\usepackage{pifont}
\usepackage[inline]{enumitem}
\usepackage{booktabs}

\usepackage{multirow}
\usepackage{fancybox} %

\renewcommand\footnotetextcopyrightpermission[1]{} %
\setcopyright{none}

\acmDOI{}

\acmISBN{}

\acmPrice{}

\usepackage{subfig}
\usepackage{xcolor}
\usepackage[inline]{enumitem}

\usepackage{times}
\usepackage[textfont={it}, font={small,bf}]{caption}

\makeatletter

\renewcommand\section{\@startsection {section}{1}{\z@}%
	{-2ex \@plus -2\p@ \@minus -.2\p@}%
	{0.5ex}
	{\normalfont\Large\bfseries}}

\renewcommand\subsection{\@startsection {subsection}{1}{\z@}%
	{-2ex \@plus -0.5ex \@minus -.5ex}%
	{0.5ex}%
	{\normalfont\large\bfseries}}

\renewcommand\subsubsection{\@startsection {subsubsection}{1}{\z@}%
    {-2ex \@plus -0.5ex \@minus -.2ex}%
    {0.5ex}%
    {\normalfont\bfseries}}

\renewcommand\paragraph{\@startsection {paragraph}{4}{\z@}%
    {-0.1ex \@plus -0.5ex \@minus -.2ex}%
    {-0.5em}%
    {\normalfont\bfseries}}

\makeatother

\definecolor{boxcolor}{gray}{0.9}
\newenvironment{colframe}{%
  \begin{Sbox}
    \begin{minipage}
      {0.96\columnwidth}
    }{%
    \end{minipage}
  \end{Sbox}
  \begin{center}
    \colorbox{boxcolor}{\TheSbox}
  \end{center}
}

\usepackage{algorithm, algorithmicx,algpseudocode}
\algtext*{EndWhile}
\algtext*{EndFor}
\algtext*{EndIf}
\algrenewcommand\algorithmicindent{0.5em}%

\newenvironment{denseitemize}{
\begin{itemize}[topsep=2pt, partopsep=0pt, leftmargin=1.5em]
  \setlength{\itemsep}{2pt}
  \setlength{\parskip}{0pt}
  \setlength{\parsep}{0pt}
}{\end{itemize}}

\newenvironment{denseenum}{
\begin{enumerate}[topsep=2pt, partopsep=0pt, leftmargin=1.5em]
  \setlength{\itemsep}{2pt}
  \setlength{\parskip}{0pt}
  \setlength{\parsep}{0pt}
}{\end{enumerate}}

\def\B4{{G-Scale}}

\def\ie{{i.e.}}
\def\eg{{e.g.}}
\def\etal{{et al.}}

\def\ninefive{$95$th}

\def\name{Terra}

\newcommand{\cmark}{\ding{51}}%
\newcommand{\xmark}{\ding{55}}%

\begin{document}
\title{\huge {\name}: Scalable Cross-Layer GDA Optimizations}

\author{Jie You}
\affiliation{%
	\institution{University of Michigan}
	\city{Ann Arbor}
	\state{Michigan}
	\postcode{48105}
}
\email{jieyou@umich.edu}

\author{Mosharaf Chowdhury}
\affiliation{%
	\institution{University of Michigan}
	\city{Ann Arbor}
	\state{Michigan}
	\postcode{48105}
}
\email{mosharaf@umich.edu}

\begin{sloppypar}
\begin{abstract}
Geo-distributed analytics (GDA) frameworks transfer large datasets over the wide-area network (WAN). 
Yet existing frameworks often ignore the WAN topology. 
This disconnect between WAN-bound applications and the WAN itself results in missed opportunities for cross-layer optimizations.

In this paper, we present {\name} to bridge this gap. 
Instead of decoupled WAN routing and GDA transfer scheduling, {\name} applies scalable cross-layer optimizations to minimize WAN transfer times for GDA jobs. 
We present a two-pronged approach:
(i) a scalable algorithm for joint routing and scheduling to make fast decisions; and 
(ii) a scalable, overlay-based enforcement mechanism that avoids expensive switch rule updates in the WAN.
Together, they enable {\name} to quickly react to WAN uncertainties such as large bandwidth fluctuations and failures in an application-aware manner as well.

Integration with the FloodLight SDN controller and Apache YARN, and evaluation on $4$ workloads and $3$ WAN topologies show that {\name} improves the average completion times of GDA jobs by $1.55\times$--$3.43\times$.
GDA jobs running with {\name} meets $2.82\times$--$4.29\times$ more deadlines and can quickly react to WAN-level events in an application-aware manner.
\end{abstract}
 
\maketitle

\section{Introduction}

To cope with the increasing number of Internet users and edge devices \cite{internet-user-growth}, large organizations leverage tens to hundreds of datacenters and edge sites \cite{azure-locs, ec2-locs, google-locs, google-dc-map} to gather data related to end-user sessions and their devices as well as monitoring logs and performance counters.
Analyzing and personalizing this data can provide tremendous value in improving user experience \cite{geode, iridium, cisco-ioe}. 
Consequently, a growing body of recent work has focused on enabling \emph{geo-distributed analytics (GDA)}, \ie, executing computation tasks on the data stored in-place at different sites instead of copying to a single datacenter.
Faster completions of these jobs can enable log processing \cite{jetstream, iridium}, SQL query processing \cite{geode, clarinet}, and machine learning \cite{gaia-cmu} over large geo-distributed datasets.
Assuming static WAN bandwidth, existing solutions optimize query planning \cite{clarinet, geode, jetstream, tetrium2018}, scheduling \cite{iridium, swag}, and algorithm design \cite{gaia-cmu} to reduce inter-datacenter transfers over the WAN.
This is because WAN bandwidth is expensive \cite{pretium, bwe} and often a major performance bottleneck for these communication-intensive jobs \cite{geode, clarinet, gaia-cmu} (\eg, due to large intermediate data transfers \cite{iridium}).

Unfortunately, existing GDA frameworks ignore the WAN topology and treat the WAN as a full mesh or a non-blocking switch \cite{clarinet, geode, jetstream, iridium, gaia-cmu, swag}.
Although these simplifications -- end-to-end tunnels and independent links, respectively -- decouple GDA systems from WAN traffic engineering, they introduce a disconnect.
Applications cannot optimize on actual points of contention which are hidden and constantly changing in the WAN; at the same time, WAN traffic engineering cannot optimize application-level objectives.

This mismatch between application- and WAN-level goals prolongs the communication stages of GDA jobs and increases their job completion times (JCT) (\S\ref{sec:motivation}).
Existing solutions that attempt to align the two \cite{rapier, mobihoc16coflow} do not scale to large WAN topologies or complex jobs, and they themselves can become the bottleneck (\S\ref{sec:idea}). 
The presence of WAN uncertainties such as large bandwidth fluctuates and link failures adds to the challenge because GDA jobs cannot rapidly adapt to changing WAN conditions.

Our goal in this paper is to speed up the communication stages of GDA jobs and to make them more robust to WAN uncertainties. 
To this end, we present {\name}, a scalable framework that bridges the gap between GDA frameworks and the WAN by co-optimizing application-level transfer scheduling and WAN-level routing decisions.
{\name}'s design is guided by two high-level observations:
\begin{denseitemize}
	\item Redundant paths in the WAN topology should be fully utilized in order to minimize GDA transfer times; and
  
	\item SD-WAN rule updates are expensive and should be avoided whenever possible for fast decision enforcement. 
\end{denseitemize}
We propose a two-pronged approach that can scale to large GDA jobs and WAN topologies while adhering to these observations. 
First, we propose a scalable algorithm to quickly compute multipath WAN routing and GDA scheduling decisions (\S\ref{sec:idea}).
To this end, we generalize existing coflow-based solutions used inside single datacenters \cite{coflow-hotnets, varys, rapier} to consider the WAN topology. 
Then we make it scalable by treating all flows between the same datacenter pair from the same coflow together instead of treating each one independently, reducing the problem size by many orders-of-magnitude.
Second, we propose a multipath overlay on top of single-path TCP connections over the entire WAN and enforce our algorithm-determined \emph{routes}, \emph{schedules}, and \emph{rates} on this overlay, limiting the need for WAN rule updates only to (re)initialization (\S\ref{sec:overview}).
Our algorithm-systems co-design can address WAN uncertainties by quickly recomputing GDA transfer schedules and reconfiguring the WAN overlay.

\begin{figure*}[!ht]
  \vspace{-0.5cm}
 \begin{tabular}[b]{cc}
   \begin{tabular}[b]{c}
     \subfloat[][MapReduce Job on WAN]{%
       \label{fig:coflow-ex-topo}%
       \includegraphics[scale=0.45]{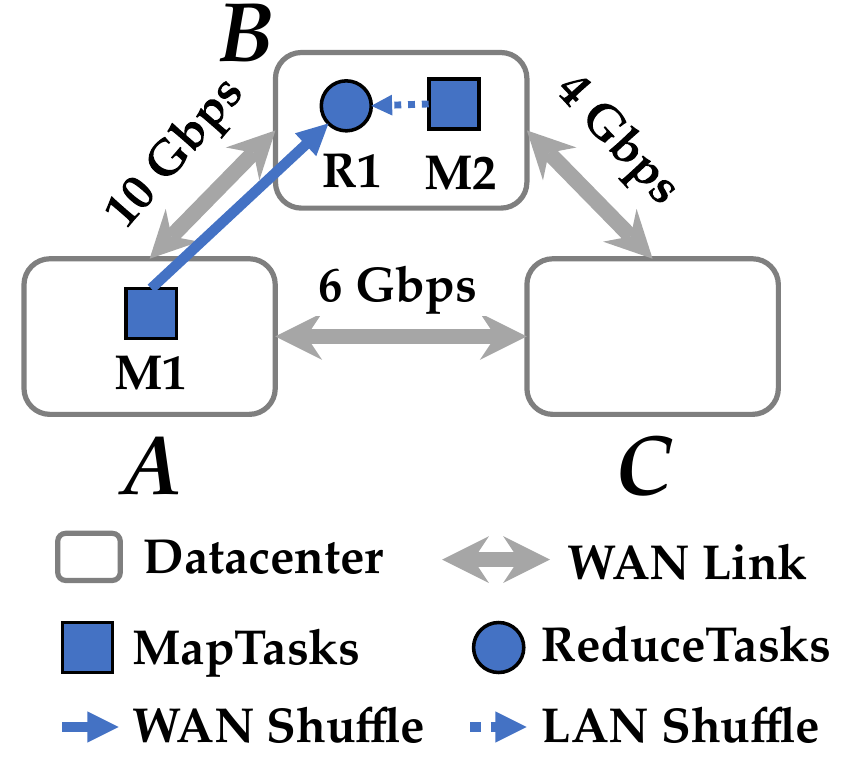}%
     }\\
     \subfloat[][WAN Traffic of 2 jobs]{%
       \label{fig:coflow-ex-setup}%
       \includegraphics[scale=0.5]{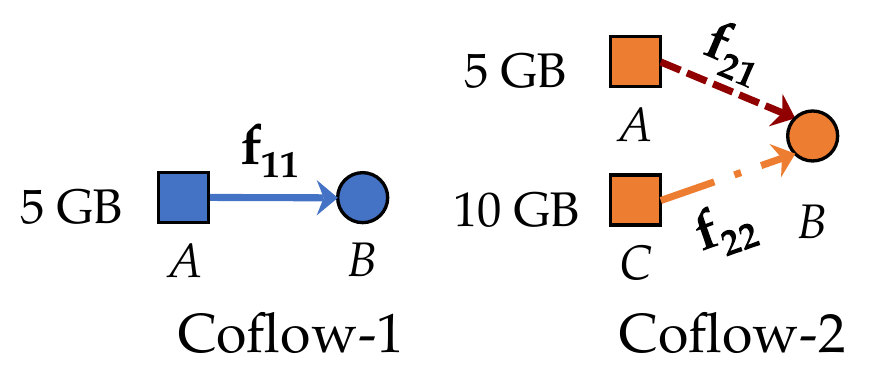}%
     }
   \end{tabular}
   &
   \begin{tabular}[b]{c}
     \subfloat[][Flow-level fair sharing]{%
       \label{fig:coflow-ex-perflow}%
       \includegraphics[scale=0.45]{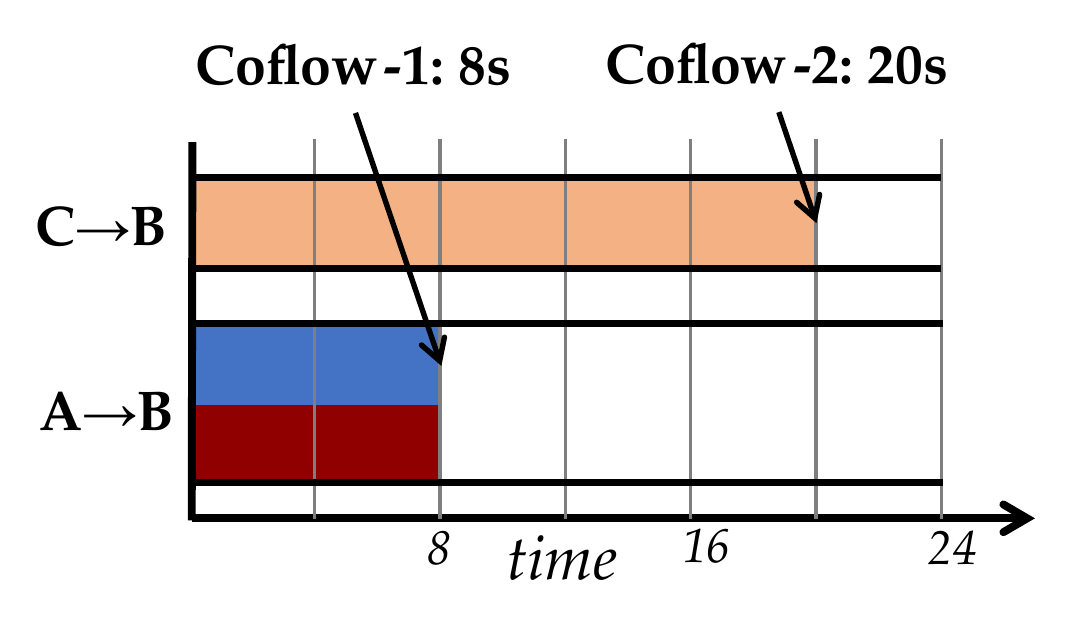}%
     }
     \hspace{1em}
     \subfloat[][Multipath]{%
       \label{fig:coflow-ex-multipath}%
       \includegraphics[scale=0.45]{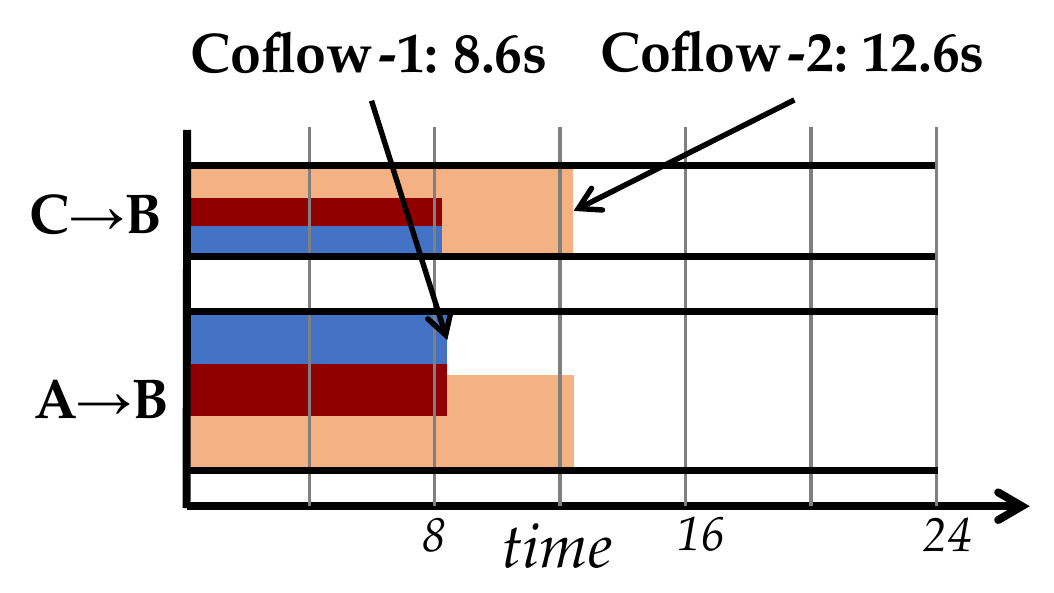}%
     }\\
     \subfloat[][Coflow scheduling]{%
       \label{fig:coflow-ex-coflow}%
       \includegraphics[scale=0.45]{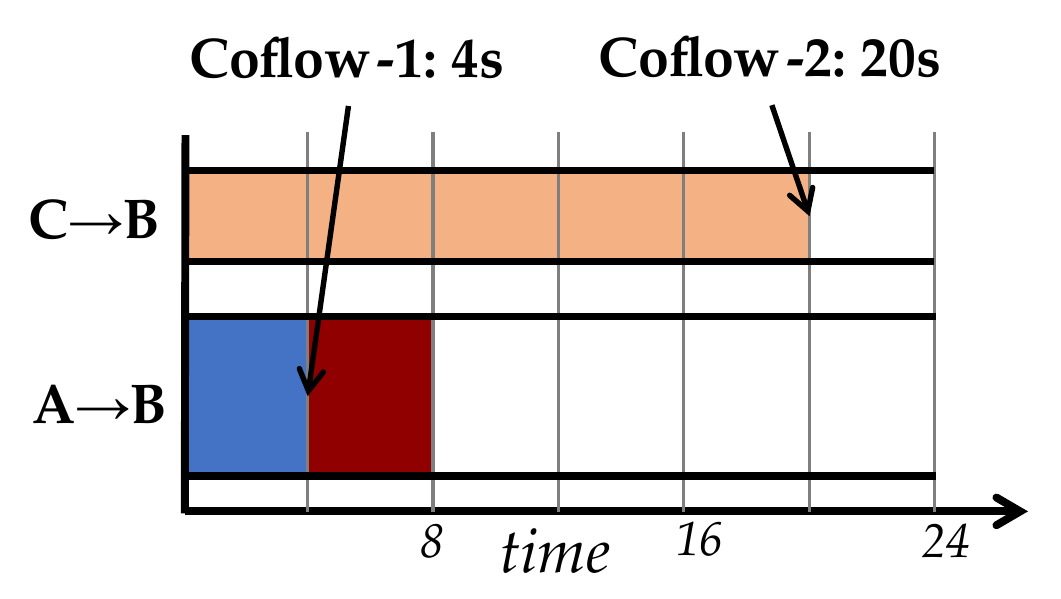}%
     }
     \hspace{1em}
     \subfloat[][{\name}]{%
       \label{fig:coflow-ex-optimal}%
       \includegraphics[scale=0.45]{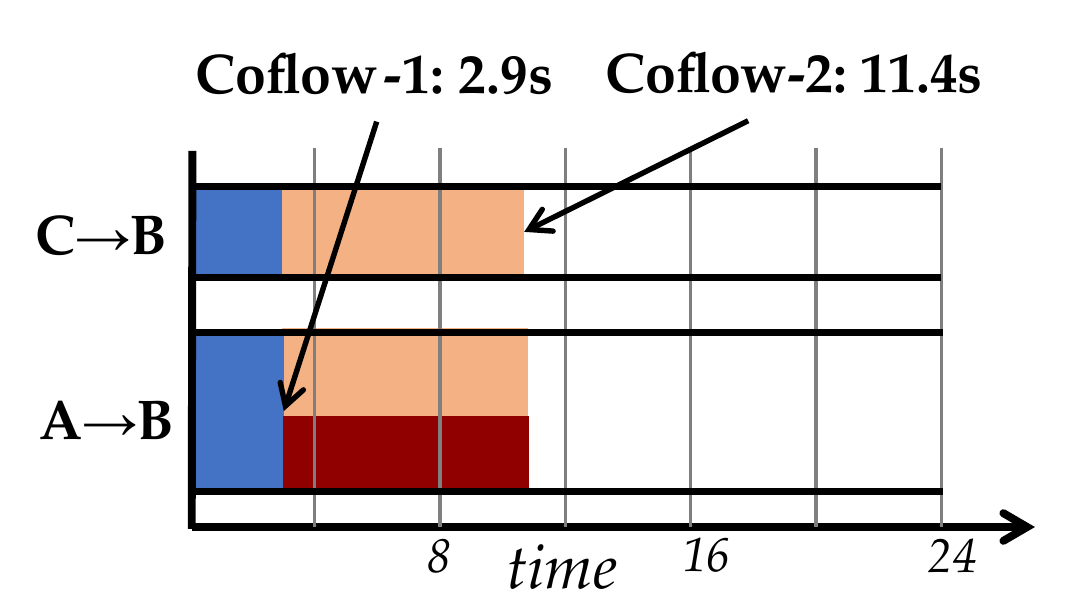}%
     }
   \end{tabular}
 \end{tabular}
  \vspace{-0.3cm}
  \caption{Opportunities for scheduling-routing co-optimization of two jobs running across three datacenters. 
    \protect\subref{fig:coflow-ex-topo} MapReduce Job running on WAN topology. 
    \protect\subref{fig:coflow-ex-setup} Coflows from Job-1 (dark/blue) has $1$ flow and Job-2 (light/orange) has $2$ flows (dark/red and light/orange).
    \protect\subref{fig:coflow-ex-perflow}--\protect\subref{fig:coflow-ex-optimal} Bandwidth allocations of the two bottleneck links ($A\rightarrow B$ and $C\rightarrow B$) w.r.t. time for $4$ different scheduling-routing policies.
    Average completion times for
    \protect\subref{fig:coflow-ex-perflow} per-flow fair sharing is $14$ seconds;
    \protect\subref{fig:coflow-ex-multipath} multipath is $10.6$ seconds;
    \protect\subref{fig:coflow-ex-coflow} intra-datacenter coflow scheduling \cite{varys, rapier} is $12$ seconds;
    \protect\subref{fig:coflow-ex-optimal} {\name} finds the optimal routing-scheduling joint solution: $7.15$ seconds.}%
  \label{fig:coflow-ex}
  \vspace{-0.4cm}
\end{figure*}
 
We have a full-stack implementation of the proposed solution (\S\ref{sec:implementation}), integrated with the FloodLight \cite{floodlight} SDN controller in the network-side and Apache YARN \cite{yarn} in the application-side.
It provides a simple API to express WAN coflows.
User-written jobs in a framework remain unmodified.

We evaluated {\name} using three WAN topologies (Microsoft's SWAN \cite{swan}, Google's {\B4} \cite{b4}, and AT\&T's MPLS topology \cite{att-mpls}) and four different workloads (BigBench \cite{bbench}, TPC-DS \cite{tpc-ds}, and TPC-H \cite{tpc-h} with complex DAGs and Facebook data transfer matrices from simple MapReduce jobs \cite{swim, coflow-benchmark}) (\S\ref{sec:eval}).
For the small-scale testbed experiment using the SWAN topology, {\name} improved the average job completion time (JCT) by $1.55\times$--$3.43\times$ on average and $2.12\times$--$8.49\times$ at the {\ninefive} percentile, while improving WAN utilization by $1.32\times$--$1.76\times$. 
For large-scale simulations, {\name} improved the average JCT by $1.04\times$--$2.53\times$ for the smallest topology (SWAN) and $1.52\times$--$26.97\times$ for the largest topology (AT\&T) against baselines such as per-flow fairness, multipath routing, SWAN-MCF \cite{swan}, Varys \cite{varys}, and Rapier \cite{rapier}.
{\name} can complete $2.82\times$--$4.29\times$ more coflows within their deadlines. 
We also show that it can react quickly to WAN events, it can scale well, and its benefits hold across a wide range of parameter and environment settings.

In summary, our contributions in this paper are three-fold:
\begin{denseenum}
	\item Identifying scalability bottlenecks in the GDA-WAN co-optimization problem both from algorithm and system design perspectives.
  
	\item A scalable algorithm that co-optimizes complex GDA jobs with large WAN topologies to minimize their data transmission times.
  
	\item A scalable system design and implementation that integrates with GDA frameworks and SD-WANs to enforce those decisions and provides large performance benefits.
	
\end{denseenum}

\section{Background and Motivation}
\label{sec:motivation}

This section provides a quick overview of GDA systems (\S\ref{sec:gda}), common WAN models used by them (\S\ref{sec:wan-model}), and the coflow abstraction (\S\ref{sec:coflow}), followed by an illustration of the advantages of application-WAN co-optimization (\S\ref{sec:benefits}).

\subsection{Geo-Distributed Analytics (GDA)}
\label{sec:gda}
GDA users submit jobs written in higher-level interfaces -- \eg, in SparkSQL \cite{sparksql}, Hive \cite{hive}, or GraphX \cite{graphx} -- typically to one central \emph{job master} that resides in one of the many distributed sites/datacenters \cite{geode, iridium}.
The master constructs an optimized execution plan \cite{clarinet} for the job and represents it as a directed acyclic graph (DAG). 
Nodes of this DAG represent computation stages with many parallel tasks and edges represent stage dependencies as well as WAN transfers between tasks in different datacenters. 
A centralized scheduler then places these tasks at machines across different datacenters based on data locality, resource availability, and their dependencies \cite{swag, geode, clarinet}.
The durations of these jobs typically range from minutes to tens of minutes and communication across the WAN is often the bottleneck \cite{iridium,clarinet}.

\begin{figure*}[!t]
  \vspace{-0.2cm}
  \centering
    \subfloat[][Two coflows]{%
      \label{fig:ft-sec2-setup}%
      \includegraphics[scale=0.5]{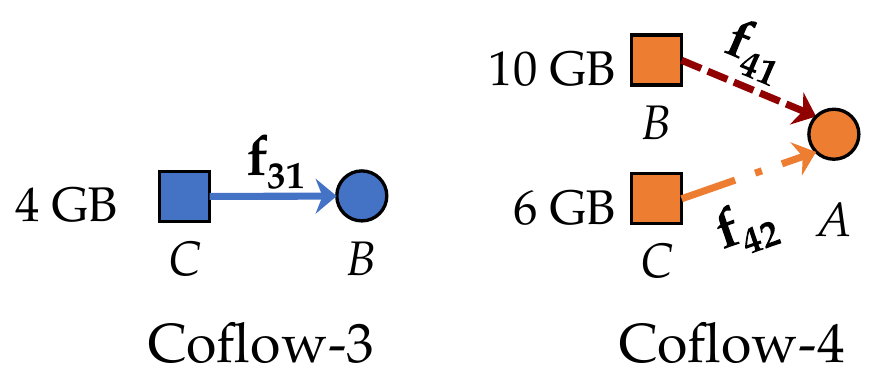}%
    }
    \hfill
    \subfloat[][Optimal]{%
      \label{fig:ft-sec2-opt}%
      \includegraphics[scale=0.45]{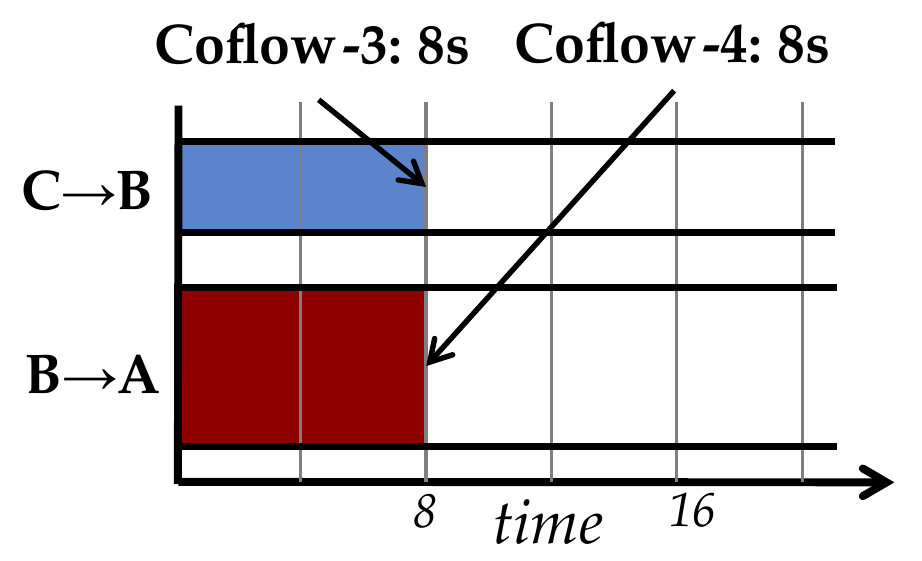}%
    }
    \hfill
    \subfloat[][Suboptimal upon failure]{%
      \label{fig:ft-sec2-failure}%
      \includegraphics[scale=0.45]{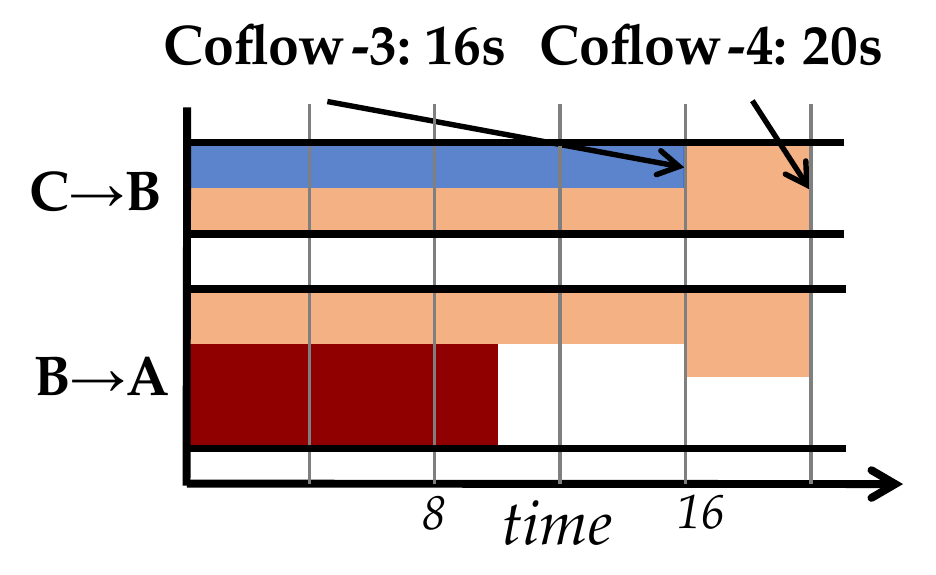}%
    }
    \hfill
    \subfloat[][Optimal after failure]{%
      \label{fig:ft-sec2-opt-failure}%
      \includegraphics[scale=0.45]{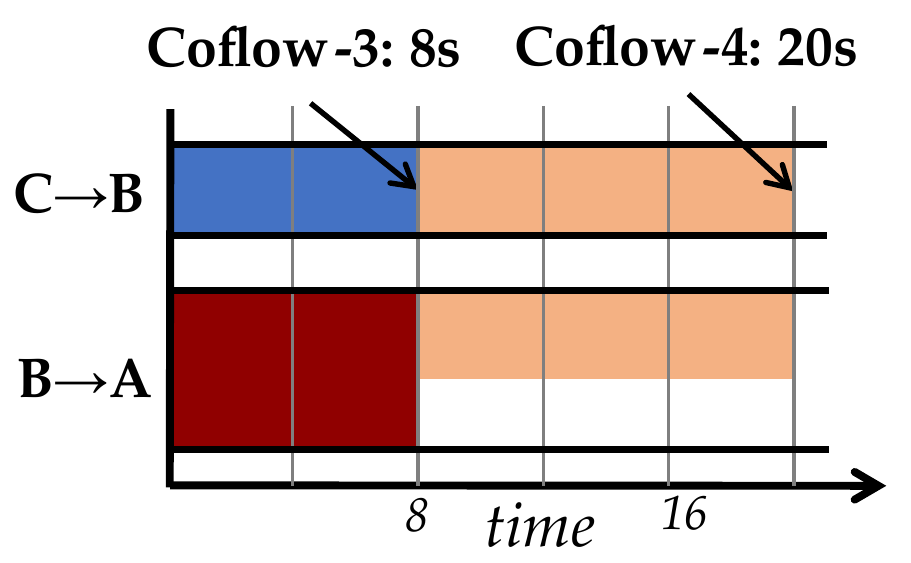}%
    }
  \vspace{-0.3cm}
  \caption{Need for application-aware WAN re-optimization. 
    \protect\subref{fig:ft-sec2-setup} On Figure~\ref{fig:coflow-ex-topo}'s topology, Coflow-3 (dark/blue) has $1$ flow and Coflow-4 (light/orange) has $2$ flows (dark/red and light/orange).
    Average completion times 
    \protect\subref{fig:ft-sec2-opt} for the optimal solution is $8$ seconds;
    \protect\subref{fig:ft-sec2-failure} after rerouting {$f_{42}$} due to the failure of the $A$--$C$ link is $18$ seconds;
    \protect\subref{fig:ft-sec2-opt-failure} for the new optimal solution after the failure is $14$ seconds.}%
  \label{fig:ft-sec2}
  \vspace{-0.4cm}
\end{figure*}

\subsection{Inter-Datacenter WAN Model}
\label{sec:wan-model}
Datacenters used by GDA frameworks are connected by a WAN. 
While such WANs have traditionally been optimized using MPLS-based \cite{mpls-te} traffic engineering, centralized SD-WANs are becoming more popular \cite{b4, swan, bwe}. 
We assume the latter, which enables {\name} to make and enforce topology-aware routing decisions.
Existing GDA systems assume either a full mesh with heterogeneous link capacities \cite{clarinet, gaia-cmu} or a non-blocking switch with contentions only at the uplinks or downlinks of each datacenter \cite{iridium}. 
As such, they miss the opportunity to utilize redundant paths in WAN.

Existing solutions assume that the WAN topology and available bandwidth remain fixed during a job's execution. 
However, this may lead to performance issues when WAN configuration is updated in the middle of a job's execution.
For example, SWAN~\cite{swan} updates WAN configurations every 5 minutes.
Because {\name} is integrated with the SD-WAN controller, unlike existing solutions, it can monitor and react to these changes.
High-priority, user-facing, and deadline-sensitive traffic are prioritized by WAN managers \cite{tempus, bwe}, including {\name}. 
So we consider a link's bandwidth to be the remaining capacity excluding those traffic.

\subsection{The Coflow Abstraction}
\label{sec:coflow}
GDA jobs typically use the same programming models (\eg, MapReduce) as traditional analytics jobs running within a single datacenter \cite{iridium, clarinet, geode}. 
These programming models often have a communication stages between computation stages, where the computation stage cannot start until all flows in the preceding communication stage have finished.
Recently, Chowdhury and Stoica defined such a collection of flows with a shared fate as a coflow \cite{coflow-hotnets}, and many have shown that minimizing a coflow's completion time (CCT) can decrease a job's completion time \cite{varys, rapier, sincronia}.
The coflow abstraction generalizes intermediate data transfers (i.e., shuffles) for GDA jobs too \cite{siphon}.

\subsection{Potential Benefit of Co-Optimization}
\label{sec:benefits}

\paragraph{Setup}
Without loss of generality, we consider a GDA job with one map stage and one reduce stage for ease of explanation. 
Now consider this job running on the WAN topology with 3 datacenters $\{A, B, C\}$ as shown in Figure~\ref{fig:coflow-ex-topo}.
Suppose a GDA query planning and task placement algorithm \cite{clarinet, geode, jetstream, tetrium2018} has put some map tasks in $A$, others in $B$, and all reduce tasks in $B$. 
Therefore, part of its shuffle traffic would be transfered over the WAN from $A$ to $B$, while the other part would be inside datacenter $B$. 
We focus on WAN traffic here because the limited bandwidth of WAN becomes the bottleneck for GDA jobs \cite{geode,iridium,clarinet}.
Assuming the total volume of intermediate data generated by $M_1$ at datacenter $A$ to be 5~GB, we can now form a coflow (Coflow-1 in Figure \ref{fig:coflow-ex-setup}).
Coflow-2 is similarly generated, but with a different communication pattern.
Our goal now is to minimize the average completion time for both coflows.

\paragraph{Existing Solutions}
First, let us consider the classic flow-level fair sharing that equally divides the $A\rightarrow B$ link between flows {$f_{11}$} and {$f_{21}$} (Figure~\ref{fig:coflow-ex-perflow}).
Thus, both flows complete by $8$ seconds, whereas {$f_{22}$} completes by $20$ seconds facing no contention.
Consequently, Coflow-1 and Coflow-2 complete in $8$ and $20$ seconds, respectively. 
The average completion time is $14$ seconds.

A simple improvement would be using multiple paths (\eg, MPTCP \cite{rfc6824}) to increase network utilization (Figure~\ref{fig:coflow-ex-multipath}). 
In this case, all the flows are split across available paths. 
Assuming equal split and fair sharing in each link, 
the average completion time is $10.6$ seconds.

Coflow-aware scheduling \cite{orchestra, varys, aalo, rapier} improves the average completion time by considering \emph{all} the flows of the same coflow together. 
In this case, {$f_{11}$} will be scheduled before {$f_{21}$} on the $A$--$B$ link (Figure~\ref{fig:coflow-ex-coflow}).
Consequently, Coflow-1 finishes in $4$ and Coflow-2 in $20$ seconds. 
The average completion time is $12$ seconds.
These coflow-based solutions still fall short either by assuming a non-blocking topology \cite{varys} or due to considering a single path \cite{rapier}.

\paragraph{Co-Optimization Finds the Optimal Solution}
So far we have shown 3 sub-optimal solutions, where they only optimize one side of the Application-WAN duo. 
However, if we consider both simultaneously and combine coflow scheduling with multipath routing together, we can achieve the optimal average completion time of only $7.15$ seconds (Figure \ref{fig:coflow-ex-optimal}).

Note that we considered only two jobs and a minimal full-mesh topology in the offline scenario. 
{\name} performs even better with more jobs, on larger WAN topologies that are not full mesh, and in online scenarios (\S\ref{sec:eval}).

\paragraph{Re-Optimization is Necessary Under Uncertainties}
Consider the same topology as in Figure~\ref{fig:coflow-ex-topo} but with two different coflows (Figure~\ref{fig:ft-sec2-setup}).
Existing WAN-agnostic solutions \cite{iridium, clarinet} will schedule Coflow-3 and Coflow-4 together to achieve the optimal average completion time of $8$ seconds.
However, if the link between $A$ and $C$ fails (or experiences a massive increase in high-priority traffic) right after the scheduling decision has been made, the WAN will reroute {$f_{42}$} and the completion times of Coflow-3 and Coflow-4 would become $16$ and $20$ seconds, respectively.
Hence, the average completion time would be $18$ seconds.

The optimal solution is rescheduling Coflow-3 before Coflow-4 so that they complete in $8$ and $20$ seconds for the new minimum average completion time of $14$ seconds.

\subsection{Summary}
Table~\ref{tab:comparison} compares the solutions discussed above across key criteria. 
The key takeaways from this section are: 
\begin{denseenum}
  \item The optimal average coflow completion time can only be achieved when jointly considering routing and scheduling;
  \item In the presence of WAN uncertainties (\eg, bandwidth fluctuations and failures), application-level scheduling must react to WAN-level routing, and vice versa.
\end{denseenum}

\newlength{\savedtabcolsep}
\setlength{\savedtabcolsep}{\tabcolsep}
\setlength\tabcolsep{2 pt}
\begin{table}[!t]
  \vspace{-0.3cm}
  \begin{center} 
  \begin{small}
    \begin{tabular}{lcccc}
    	\hline
    	\hline
        &WAN-     &Leverages   &App-  &Re- \\
        &Aware*   &Multipath &Aware &Optimizes$\dagger$\\
      \hline
      Per-Flow/TCP    & \xmark & \xmark & \xmark & \xmark \\
      Multipath/MPTCP       & \xmark & \cmark & \xmark & \xmark \\
      Datacenter Coflows    & \xmark & \xmark & \cmark & \xmark	\\
      SD-WANs         & \cmark & \cmark & \xmark & \xmark	\\
      GDA Systems     & \xmark & \xmark & \cmark & \xmark	\\
      Terra  & \cmark & \cmark & \cmark & \cmark	\\
      \hline \hline
    \end{tabular}
  \end{small}
  \end{center}
  \caption{{\name} vs existing solutions. 
    *WAN-Aware: does not assume full-mesh topology, non-blocking core, or symmetric paths. 
    $\dagger$Re-Optimize: application-aware re-scheduling and rerouting of WAN transfers.}
  \label{tab:comparison}
  \vspace{-0.5cm}
\end{table}
\setlength\tabcolsep{\savedtabcolsep}

\section{{\name}: Algorithm Design}
\label{sec:idea}

Given the benefits of co-optimization, the need for fast re-optimizations, and the scale and heterogeneity of the WAN, we must design a cross-layer solution that can perform well at WAN scale. 
This requires both designing a scalable algorithmic solution that can quickly make joint scheduling and routing decisions and a scalable system design that can quickly enforce those decisions throughout the WAN.
In this section, we focus on the former. 
Section~\ref{sec:overview} discusses the latter.

\subsection{Minimizing the Average Completion Time}
\label{sec:mincct}
{\name}'s primary goal is faster completions of WAN transfers from geo-distributed jobs, \ie, \emph{minimizing the average CCT}.
Given a coflow, {\name} must decide \emph{when to start} individual flows, \emph{at what rate} to send them, and \emph{which path(s)} each flow should take.

This problem is computationally intractable even when all coflows start together, their traffic matrices are known a priori, and the WAN is non-blocking.
Because inter-coflow scheduling in a non-blocking datacenter is known to be NP-hard under these assumptions \cite{varys}, the counterpart on a general WAN topology is NP-hard too.
Given the intractability of the problem, we first focus on designing an efficient offline heuristic (\S\ref{sec:one-job}--\S\ref{sec:idea-multicoflow}), then extend to online scenarios (\S\ref{sec:online}).

Consider a WAN graph $G=(V,E)$, where $V$ is the set of nodes -- sets of datacenters -- and $E$ is the set of WAN links.
We represent multiple physical links between $u$ and $v$ ($u,v \in V$) with one logical link $e=(u,v) \in E$ with the cumulative bandwidth.
At time $T$, $e$'s available bandwidth is $c_T(u,w)$.
We consider the offline problem of scheduling-routing $|\mathbb{C}|$ GDA coflows ($\mathbb{C} = \{C_1, C_2, \ldots, C_{|\mathbb{C}|}\}$) that arrived at time $T$.

\begin{figure}[!t]
\vspace{-0.1cm}
	\centering
	\includegraphics[width=0.6\linewidth]{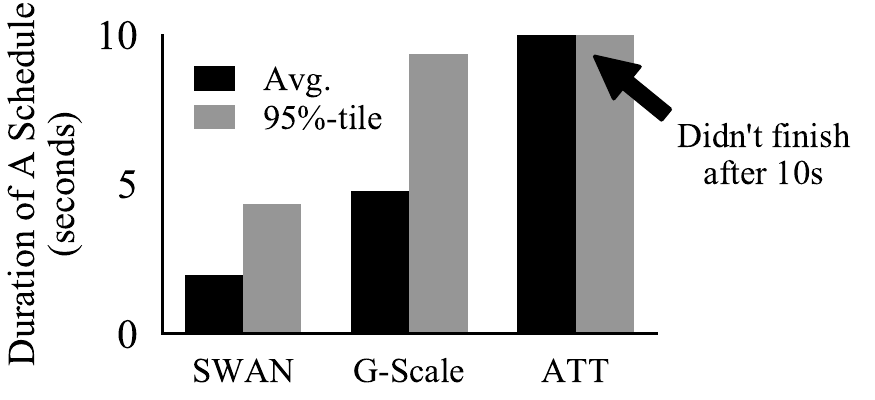}%
	\vspace{-0.3cm}
	\caption{Scheduling overhead of a state-of-the-art solution \cite{rapier}. }
	\label{fig:rapier-overhead}
\end{figure}

\begin{figure}[!t]
	\centering
	\vspace{-0.5cm}
	\hfill
	\subfloat[][Coflow with $16n$ flows.]{%
		\label{fig:flowgroup-coflow3}%
		\hspace{5mm}
		\includegraphics[width=0.35\linewidth]{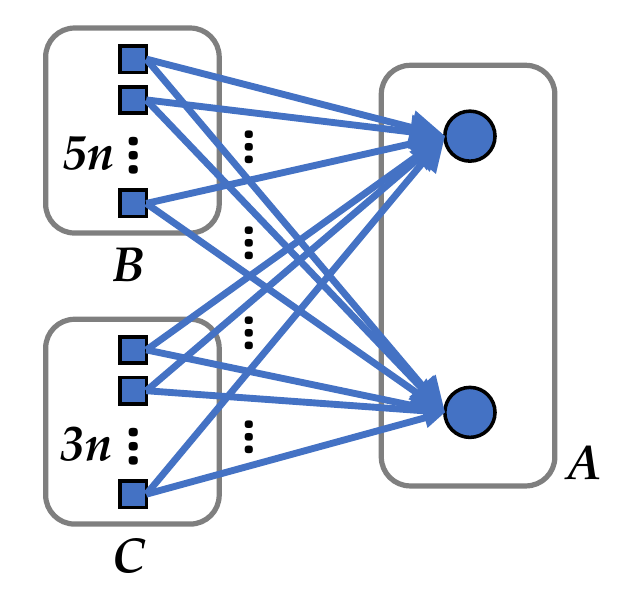}%
		\hspace{5mm}
	}
	\hfill
	\subfloat[][FlowGroups of this Coflow.]{%
		\label{fig:flowgroup-coflow3-2}%
		\hspace{5mm}
		\includegraphics[width=0.35\linewidth]{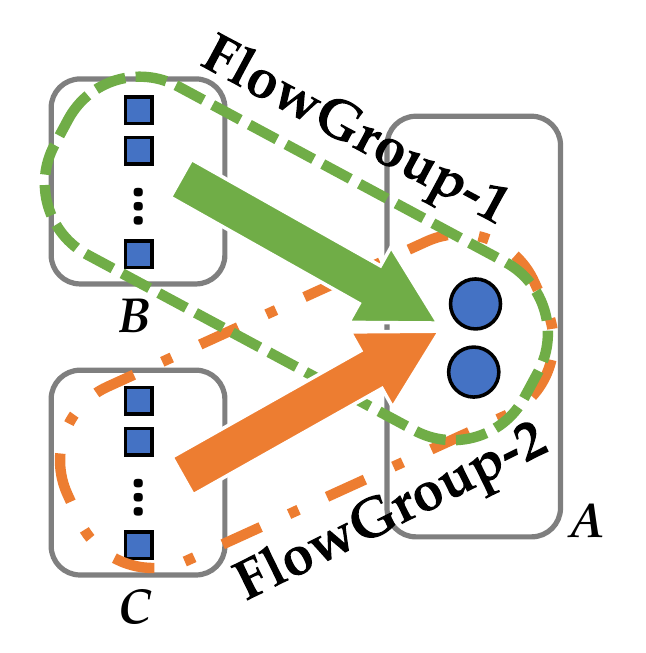}%
		\hspace{5mm}
	}
	\hfill
	\vspace{-0.2cm}
	\caption{Scaling down the number of flows in a coflow. }
	\label{fig:flowgroup}
\end{figure}

\subsubsection{The Minimum CCT of a Single Coflow}
\label{sec:one-job}
We start by focusing on a single coflow and decide \emph{how to route} its flows and \emph{at what rate} to send the traffic. 

\paragraph{Scalability Limitation}
Calculating the rate and routing for \emph{every} single flow is impractical, because the number of flows for even one coflow can be very large.
Per our measurement, the computation time of a state-of-the-art solution that considers coflow routing \cite{rapier} is $1.952$ seconds on average for the BigBench workload on the SWAN topology. 
The computation overhead only increases for even larger topologies (Figure~\ref{fig:rapier-overhead}).
Clearly, we cannot calculate rate allocation and routing for each flow of each coflow.

\paragraph{Per-Flow Rate Allocation is Unnecessary}
Existing solutions show minimal coflow completion time can be achieved by enforcing per-flow rate to ensure that \emph{all of its flows finish together} \cite{varys, orchestra, rapier}. 
However, we observe that we can still achieve minimal coflow completion time, even when individual flows do not finish together. 

Consider a MapReduce job running on the same WAN provided in Figure~\ref{fig:coflow-ex-topo}. 
Assume there are $5n$ map tasks placed in $B$, $3n$ map tasks placed in $C$ and 2 reduce tasks placed in $A$ (Figure \ref{fig:flowgroup-coflow3}). 
So there are a total of $16n$ flows in this coflow.
Suppose for each flow we need to send 1~GB data, enforcing \emph{all flows finish together} gives all flows $1/n$~Gbps throughput.
This allocation gives a minimal coflow completion time of $8n$ seconds, and both link $B\rightarrow A$ and $C\rightarrow A$ are fully utilized all the time.

Now, we take all the flows traversing through link $B\rightarrow A$, and change their rate allocation -- we schedule them one-at-a-time in the FIFO order, allocating the entire bandwidth of $B\rightarrow A$ (10~Gbps) to each of them.
We can still achieve the same CCT of $8n$ seconds.
This gives us the following lemma.

\begin{lemma}
\label{lemma:fg}
If multiple flows of the same coflow have the same <$src\_datacenter, dst\_datacenter$> pair, all work-conserving rate allocation of them will achieve the same completion time.
\end{lemma}

Consequently, we can group flows within a coflow by their <$src\_datacenter, dst\_datacenter$> tuple. 
The rates of individual flows within such a group do not directly affect the coflow completion time, as long as the total rate of the group remains the same. 
This grouping is similar to that of FlowGroup \cite{bwe}; for simplicity, we refer to such groups of flows as FlowGroups. 

The notion of FlowGroup brings performance improvements in both calculating and enforcing the rate allocation. 
Because we only need per-FlowGroup rate allocation, the scale of our problem formulation is reduced to orders-of-magnitude smaller ($O(|FlowGroups|/|Flows|)$). 
For example, in Figure~\ref{fig:flowgroup-coflow3-2}, $16n$ flows become only 2 FlowGroups.
This significantly lowers our scheduling overhead (\S\ref{sec:eval-scalability}).

\paragraph{Solution Approach}
We can now formulate an optimization problem to minimize the CCT for a single coflow on a general topology. 
Previous works \cite{rapier} assumed that a flow can only traverse through a single path, leading to an Integer Linear Programming (ILP) formulation, which is computation-intensive.
Because of Lemma~\ref{lemma:fg}, we can assume that a FlowGroup can be split across many paths, therefore eliminating all integral constraints and leading to a LP formulation.

We organize our solution in two steps:
\begin{denseenum}
  \item Scale down by coalescing flows into FlowGroups; and
  \item Obtain fractional routes for FlowGroups while minimizing the CCT. 
\end{denseenum}

\paragraph{Step 1: Scaling Down Using FlowGroups}
In this step, we collapse all flows from the \emph{same} coflow with the same <$src\_datacenter, dst\_datacenter$> tuple to one FlowGroup. 
We can now represent a coflow $C_i$ as a collection of FlowGroups $\mathbf{D}_i=[d_i(u,v)]_{|D| \times |D|}$, where $|d_i(u, v)|$ represents the total amount WAN transfers between the machines of $C_i$ in datacenters $u$ and $v$.
$\mathbb{D}_i$ represents the set of FlowGroups with non-zero volumes in $\mathbf{D}_i$.

\paragraph{Step 2: Determining CCT Lower-Bound}
We now determine the \emph{paths} and \emph{rates} of individual FlowGroups to minimize the CCT.
We denote the completion time of coflow $C_i$ by $\Gamma_i$.
Here $\Gamma_i$ is defined as:
\[
\Gamma_i = \max_k T(d^k_i), ~~~~ d^k_i \in \mathbb{D}_i,
\]
where $d^k_i$ is the k-th FlowGroup of $C_i$, and $T(\cdot)$ is the completion time of a FlowGroup.
Hence, the slowest FlowGroup $d^k_i \in \mathbb{D}_i$ determines $\Gamma_i$. Our objective is given as:
\begin{equation}
  \text{Minimize } \Gamma_i\label{eq:goal}
\end{equation}

Let us represent the bandwidth allocation of the $k$-th FlowGroup in $\mathbb{D}_i$ ($1 \leq k \leq |\mathbb{D}_i|$) with size $|d^k_i|$ between nodes $u$ and $v$ by $f^k(u,v)$ where $u,v \in V$. 
To minimize $\Gamma_i$, we generalize WSS \cite{orchestra} and MADD \cite{varys} to multiple paths to enforce \emph{equal rate of progress} for all FlowGroups. 
For each FlowGroup, we can then ensure that they make $1/\Gamma_i$ progress every time unit by enforcing the following constraints:
\begin{gather*}
	\sum_{w \in V} f^k(\mbox{src}(d^k_i), w) =  |d^k_i| / \Gamma_i
	\label{eq:sourcedemand} \\
	\sum_{w \in V} f^k(w, \mbox{dst}(d^k_i))  =  |d^k_i| / \Gamma_i
	\label{eq:sinkdemand}
\end{gather*}
The former ensures that the outgoing rate of a FlowGroup is proportional to its volume, while the latter enforces the same on the receiving end.
Finally, we enforce usual capacity and flow conservation as follows.
\begin{gather*}
	\sum_{v \in V} f^k(u, v) + f^k(v, u) = 0, \quad\forall u \neq \mbox{src}(d^k_i), \mbox{dst}(d^k_i)
	\label{eq:flowconservation} \\
	\sum f^k(u,v) \leq c_T(u,v)
	\label{eq:bandwithcapacity} \\
	f^k(u,v) \geq 0 
	\label{eq:positive}
\end{gather*}

Note that enforcing $1/\Gamma_i$ rate to all FlowGroups leaves the maximum amount of bandwidth possible for other coflows that are scheduled after $C_i$ without sacrificing $C_i$'s CCT. 
Work conservation uses up any remaining bandwidth (\S~\ref{sec:idea-multicoflow}).

If Optimization~\eqref{eq:goal} has a feasible solution for $C_i$, it creates a matrix $\mathbf{f}_i^k=[f^k(u, v)]_{|V| \times |V|}$ for each FlowGroup corresponding to its allocations on individual links of the WAN.
Because $f^k(u,v)$ can be non-integers, a FlowGroup can be subdivided across many paths from $u$ to $v$.
We enforce this using an overlay in our systems design (\S\ref{sec:overview}).

\floatname{algorithm}{Pseudocode}
\begin{algorithm}[t!]
	
	\begin{small}
		\begin{algorithmic}[1]
			
			\Procedure{allocBandwidth}{Coflows $\mathbb{C}$, WAN $G$}
			\State{Scale down $G$ by ($1-\alpha)$}\label{line:nostarve} \Comment{Starvation freedom}
			\State{$\mathbb{C}_{\mbox{Failed}} = \emptyset$} \Comment{Coflows not scheduled in entirety}
			\ForAll{$C_i \in \mathbb{C}$}
			\State{$\Gamma_i, \mathbf{f}_i^k$ = Solve Optimization~\eqref{eq:goal} for $C_i$ on $G$}
			\If{$\Gamma_i = -1$} %
			\State{$\mathbb{C}_{\mbox{Failed}} = \mathbb{C}_{\mbox{Failed}} \bigcup C_i$}
			\State{{\bf continue~}}
			\EndIf
			\If{$D_i \neq -1$} \Comment{$C_i$ has a deadline}
			\State{Scale down $\mathbf{f}_i^k$ by $\Gamma_i/D_i$}
			\EndIf    
			\State{$\mathbb{P}_i^k$ = \mbox{\{End-to-end paths from $\mathbf{f}_i^k$ allocations\}}}
			\State{$G$ = Updated $G$ by subtracting $\mathbf{f}_i^k$ allocations}
			\EndFor
			\State{$C^{*} = \bigcup C$ for all ${C \in \mathbb{C}_{\mbox{Failed}}}$}
			\State{Allocate $C^{*}$ on $G$ using MCF}\label{line:conserve} \Comment{Work conservation}
			\State{Allocate $\mathbb{C} \setminus C^{*}$ on $G$ using MCF}\label{line:conserve2} 
			\EndProcedure
			
			\Statex
			
			\Procedure{minimizeCCTOffline}{Coflows $\mathbb{C}$}
			\State{$\mathbb{C}'$ = Sort $\mathbb{C}$ by increasing $\Gamma_i$}
			\State{allocBandwidth($\mathbb{C}'$, $G$)}
			\EndProcedure
			
		\end{algorithmic}
	\end{small}
	\caption{Offline Scheduling-Routing}
	\label{alg:mif-offline}
\end{algorithm}

\subsubsection{Scheduling Multiple Coflows}
\label{sec:idea-multicoflow}
We now move on to considering multiple coflows in the offline scenario.
Given multiple coflows, scheduling one coflow can impact the CCTs of all other coflows scheduled afterward. 
Consequently, a natural extension of the SRTF policy is sorting the coflows by their $\Gamma$ values and scheduling them in that order ({\sc minimizeCCTOffline} in Pseudocode~\ref{alg:mif-offline}). 
This requires solving $O(N)$ instances of Optimization~\eqref{eq:goal} during each scheduling round, which is activated by a coflow's arrival, completion, and WAN events. 
We schedule a coflow if all of its FlowGroups can be scheduled simultaneously. 

\paragraph{Work Conservation}
If the WAN is still not fully utilized after scheduling all coflows that can be scheduled in their entirety, we run a max-min multi-commodity flow (MCF) formulation similar to \cite{swan} on a combination of coflows (prioritizing $\mathbb{C}_{\mbox{Failed}}$) to ensure work conservation and maximize WAN utilization (line \ref{line:conserve},\ref{line:conserve2} in Pseudocode~\ref{alg:mif-offline}).

\floatname{algorithm}{Pseudocode}
\begin{algorithm}[t!]

\begin{small}
\begin{algorithmic}[1]

\Procedure{OnArrival}{Coflows $\mathbb{C}$, Coflow $C_i$}
  \If{$D_i \neq -1$}\Comment{$C_i$ has a deadline}
    \State{$G'$ = Scale down $G$ by ($1-\alpha)$}
    \State{$G'$ = $G'$ - $\{\mathbf{f}_j^k\} \; \forall \; \mbox{admitted} \; C_j$} \Comment{Guarantee admitted}
    \State{$\Gamma_i$ = Solve Optimization~\eqref{eq:goal} for $C_i$ on $G'$} 
    \If{$\Gamma_i > \eta D_i$}
      \State{{ Reject $C_i$~}}\label{line:admission} \Comment{Reject $C_i$ if its deadline cannot be met}
    \EndIf
  \EndIf
  \State{$\mathbb{C}$ = $\mathbb{C} \bigcup C_i$}
  \State{$\mathbb{C}'$ = Sort $\mathbb{C}$ by decreasing $D_i$ and then by increasing $\Gamma_i$}
  \State{allocBandwidth($\mathbb{J}'$, $G$)}
\EndProcedure

\end{algorithmic}
\end{small}
\caption{Online Scheduling-Routing}
\label{alg:mif-online}
\end{algorithm}

\subsubsection{From Offline to Online}
\label{sec:online}
So far, we have assumed that all coflows arrive together and are known a priori. 
However, in practice, coflows arrive over time as DAG dependencies are met.
Additionally, WAN links can fail and its bandwidth can fluctuate.
Scheduling coflows in the FIFO order \cite{baraat, orchestra} is a simple solution, but it can result in head-of-line blocking \cite{pdq, pfabric, varys}. 
Instead, preemption can minimize the average completion time \cite{pdq, pfabric, varys}.

\paragraph{Starvation-Free Preemption}
We allow coflows with smaller remaining completion time to preempt larger ones to avoid head-of-line blocking (Pseudocode~\ref{alg:mif-online}).
To avoid starvation issue that may arise with preemptive scheduling, we guarantee each coflow to receive some share of the network -- specifically, $\alpha$ fraction of the WAN capacity is shared between preempted coflows (line \ref{line:nostarve} in Pseudocode~\ref{alg:mif-offline}). 
By default, $\alpha=0.1$.

\paragraph{Scalable Online Scheduling}
In the online scenario, many events that trigger re-optimization may arrive at arbitrary time:
\begin{denseenum}
	\item Coflow being submitted as dependencies are met;
	\item FlowGroup finishes;
	\item Coflow finishes because all its FlowGroups finished;
	\item WAN topology changes because of bandwidth fluctuations/failures.
\end{denseenum}

Running the offline algorithm upon each event would cause high complexity.
{\name} avoids this high complexity by categorizing the events, and only re-optimizing those FlowGroups that need update.
For WAN bandwidth fluctuations, we consider $\rho=25\%$ to be the threshold for significant bandwidth change that can cause a rescheduling, filtering out short-term fluctuations. 

\subsection{Extensions}
\label{sec:extensions}

\paragraph{Supporting Deadlines}
To provide guaranteed completion of a coflow $C_i$ within its deadline ($D_i$), {\name} uses admission control.
We admit a coflow, if it can meet its deadline without violating that of any other already-admitted coflow's deadline -- \ie, if its deadline is not further from its minimum completion time ($\Gamma_i$) in the current WAN condition (line \ref{line:admission} in Pseudocode~\ref{alg:mif-online}). 
Note that we use a relaxation factor $\eta$ ($\eta > 1$) to mitigate the variability of WAN. 
However, when the bandwidth fluctuation is more than $(\eta-1)$, no deadlines can be guaranteed.
An admitted coflow is never preempted. %

Completing a coflow faster than its deadline has no benefit \cite{varys}.
Hence, a known optimization is elongating its CCT until the deadline and sharing the remaining bandwidth with others. 
This can be done by scaling the $f^k(u,v)$ values by $\Gamma_i/D_i$.

\paragraph{Supporting DAGs and Pipelined Workloads}
Many data analytics jobs form DAGs of computation stages with communication stages or coflows in between \cite{tez, dryad, spark, hive, sparksql}. 
Job masters can submit requests for each coflow in a DAG independently to {\name} as dependencies are met. 
Job masters can also submit a coflow with only some of its flows as soon as their dependencies are met, and then update the coflow to add more flows if more dependencies are satisfied. 
This is useful when the preceding computation tasks do not finish at the same time. 
In this case, {\name} tries to finish all the submitted flows of the coflow together, eventually finishing all the flows together.
Our evaluation shows that this simple strategy performs well (\S\ref{sec:eval}). 
Although it may be possible to perform DAG-aware optimizations \cite{carbyne, graphene}, we consider that to be a job master-specific decision and out of {\name}'s purview.

\section{{\name}: System Design}
\label{sec:overview}
So far we have focused on designing a scalable algorithm for minimizing the average coflow completion time (\S\ref{sec:idea}).
In this section, we discuss how to implement the solution in a scalable manner too. 
Furthermore, we consider how to make it robust to WAN variabilities.
We start with an architectural overview of the whole system and then provide insights into designing individual components.

\begin{figure}[!t]
	\centering
	\includegraphics[width=0.7\linewidth]{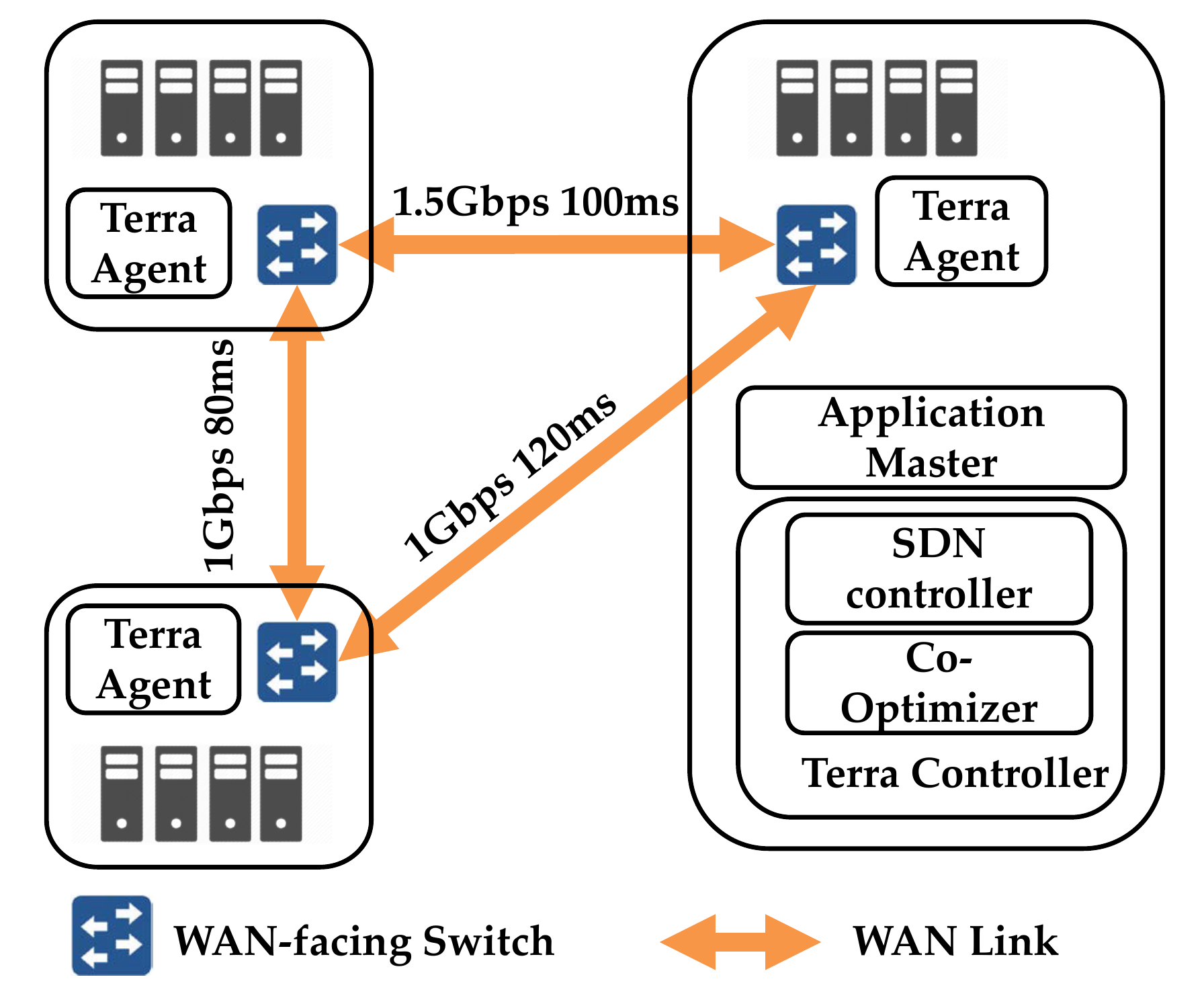}%
	\vspace{-0.3cm}
	\caption{{\name} architecture. 
		GDA jobs interact with {\name} using a client library.
		{\name} controller leverages an SD-WAN controller to make scheduling-routing co-optimization decisions that are enforced through {\name} agents (only one agent is shown in figure). }
	\label{fig:architecture}
\end{figure}

\subsection{Architectural Overview}
\label{sec:arch-overview}
As shown in Figure~\ref{fig:architecture}, {\name} has two primary components. 
A logically centralized \emph{{\name} controller} orchestrates all data transfers.  
In each datacenter, a set of \emph{{\name} agents} coordinate with the controller and transfer data on behalf of the jobs. 
Interactions between them can be summarized as follows:
\begin{denseenum}
  \item Job master(s) submit coflows to the {\name} controller using the {\name} API (\S\ref{sec:api}).
    A job can submit multiple coflows as their dependencies are met. 

  \item The controller maintains an up-to-date, global view of the WAN and coflows. 
    Given these information, it computes which jobs to schedule, which WAN paths to use, and at what rates data should be sent via specific paths (\S\ref{sec:idea}).
      
  \item Finally, the controller sends path and rate information to corresponding {\name} agents, which perform rate limiting across multiple paths (\S\ref{sec:impl-wan}). 
\end{denseenum}
The entire process takes place in an online manner. 

\subsection{Why Use a Centralized Design?}
In {\name}, all scheduling and routing decisions are made by its centralized master.
This is because \emph{making such decisions without global knowledge can lead to arbitrarily worse performance}.
Prior work \cite{aalo} has already shown that when a coflow scheduler does not coordinate, the average coflow completion time for $n$ coflows has an approximation ratio lower-bound of $\Omega(\sqrt{n})$ -- \ie, it keeps becoming worse with increasing $n$.
Because {\name}'s problem formulation generalizes the datacenter coflow scheduling problem -- specifically, instead of considering network bottlenecks only at the edge, we consider possible bottlenecks anywhere in the WAN -- the worst-case lower-bound in this scenario is at least as bad.
As such, we must have coordination among all the components of {\name} over the entire WAN.

Naturally, {\name}'s centralized design brings scalability concerns to its design forefront both during its normal operations, where every arrival or departure of a coflow calls for rescheduling, and in the presence of uncertainties such as WAN bandwidth fluctuations and failures that also require rescheduling. 
Each rescheduling round requires a central computation, followed by the dissemination and enforcement of central decisions.
We discussed the scalability aspects of the former in Section~\ref{sec:idea}, and we discuss the latter in the following.

\subsection{Scalability}
\label{sec:sys-scal}

\paragraph{Minimizing Scheduling Overhead}
Because {\name} must consider routing, it cannot use existing topology-agnostic heuristics \cite{varys, aalo}. 
However, the total number of flows in a GDA coflow adds significant time complexity to an integer linear program-based solution. 
Consequently, {\name} leverages the FlowGroup abstraction \cite{bwe} that allows us to remove the integral constraints, leading to a practical solution (\S\ref{sec:idea}). 
Each scheduling round takes O(100) milliseconds for topologies described in \cite{swan} and \cite{b4}, and O(1) seconds for larger topologies with O(10) datacenters (\S\ref{sec:eval-scalability}).
Given that many GDA jobs take several minutes to complete \cite{clarinet}, {\name} is not time-constrained in decision making and dissemination. 
Finally, because most traffic come from large jobs \cite{pacman, varys, dolly, clarinet}, {\name} can allow sub-second coflows -- \ie, only a few RTTs over the WAN -- to proceed without any coordination. 
This is similar to how SD-WANs handle interactive services \cite{swan, b4}. 

\paragraph{Restricting the Number of Paths ($k$)}
As explained in prior work \cite{swan}, running an unconstrained MCF instance may result in allocations that require many rules in switches. 
Constraining the number of paths for each FlowGroup can mitigate these issues, but it may lead to suboptimal overall WAN utilization. 
Although using $15$ shortest paths between each pair of datacenters worked well in that prior work, it can vary between WANs and must be determined experimentally by the operator. 
Operators can set any (sets of) $f^k(u,v)=0$ to enforce such constraints in {\name} (\S\ref{sec:eval-sensitivity}).
$k$ also dictates how many connections {\name} must maintain between agent pairs.

\paragraph{Minimizing Rule Updates in the WAN}
In the context of {\name}, an additional challenge is minimizing expensive rule updates throughout the entire WAN caused by route changes \cite{swan}.
Instead of setting up new rules for each flow \cite{swan}, {\name} maintains a set of single-path persistent flows between each datacenter pairs and sets up only one set of rules for each persistent flow. 
The controller sets up forwarding rules in the SD-WAN to enforce their paths.
Directing a data transfer through a specific path is then simply reusing corresponding pre-established flows. 
To completely avoid expensive rule updates, {\name} reuses persistent connections for each of the $k$ paths between two datacenters and performs communication on behalf of the applications (\S\ref{sec:impl-wan}). 
WAN states change only when these flows are initialized or reestablished after failures.
A collateral benefit of this approach is that {\name} uses only a small number of rules in each switch -- \eg, up to 168 in a switch for the SWAN~\cite{swan} topology in our evaluation.

\subsection{Robustness to Uncertainties}
Failures of {\name} agents do not permanently affect job executions because frameworks can fall back to default transfer mechanisms. 
States in {\name} agents can be rebuilt upon restart when they contact the controller. 
Failure of the {\name} controller is tied to that of the SDN controller.
Its states can be rebuilt after all the {\name} agents and job masters reconnect. 

{\name} is robust to WAN events such as link or switch failures and large bandwidth fluctuations from background traffic, because it can observe and reschedule according to the latest WAN state. 
Upon such events, {\name} updates its internal WAN representation, recomputes the set of viable paths between datacenter pairs, and updates corresponding schedules.
The entire process takes a small fraction of a typical GDA job's duration (\S\ref{sec:eval-failure}). 

Because data transfers are decoupled from the jobs, we do not have to consider sender (Map) task failures. 
If a receiver is restarted in another machine of the same datacenter upon failure, {\name} API supports updating destination(s) of a submitted coflow. 
\section{Implementation Details}
\label{sec:implementation}
We have implemented {\name} in about 8000 lines of Java including integrations with the Floodlight \cite{floodlight} SDN controller and Apache YARN \cite{yarn}.
In this section, we discuss {\name}'s integration with the SD-WAN and existing GDA frameworks.

\subsection{Integration with SD-WAN}
\label{sec:impl-wan}
WAN topologies have many redundant paths, and {\name} utilizes them to minimize coflow completion times and maximize WAN utilization.
Enforcing the multipath rate allocations of {\name}'s optimizer poses several practical challenges.

\paragraph{Multipath Data Transfers via Application-Layer Overlay}
The first challenge is emulating a multipath data transfer layer on top of single-path transport protocols such as TCP.
{\name} creates an application-layer overlay network over the WAN using persistent connections between agents.
For each path between each agent pair, one or more persistent TCP connections are maintained.
All data transfers happen over these pre-established connections.
The controller establishes these routes throughout the WAN during an offline initialization phase. 
Then the TCP connections are reused over the entire runtime for multiple coflows.

\paragraph{Per-FlowGroup Rate Enforcement}
The second challenge is enforcing rate allocation for each FlowGroup.
{\name} maps a FlowGroup across multiple end-to-end TCP flows, each with a unique combination of a sending agent, a receiving agent, its sending rate, and the path.
When a FlowGroup is scheduled to start, be preempted, or change rate, the controller informs its sending agents, which transmit data through pre-established connections for each path at designated rates.

\paragraph{Handling WAN Latency Heterogeneity}
Emulating multipath transmission over TCP flows brings a new challenge. 
Because of heterogeneous latencies between datacenter pairs, such multipath transmissions can incur many out-of-order data chunks on the receiver side.
{\name} buffers any out-of-order data to a block device and provides only in-order data to GDA jobs, mitigating this issue.

\paragraph{Supporting Other Transport Layer}
Our current implementation is based upon TCP. 
There is also a possible way forward to implement {\name} on top of MPTCP \cite{mptcp-paper} by enforcing {\name}-calculated rates directly via MPTCP's subflow scheduler.

\subsection{Integration with GDA Framework}
\paragraph{{\name} Interface}
\label{sec:api}
{\name} provides a simple API for job masters to submit new coflows, check their status, and update submitted coflows. 
A job master can submit a coflow transfer request with the set of its flows and an optional deadline, and it receives a unique CoflowId (-1 if the coflow has a deadline that cannot be met).
{
\begin{colframe}
	\fontsize{9.5pt}{11.4pt}
\begin{Verbatim}
val cId = submitCoflow(Flows, [deadline])
\end{Verbatim}
\end{colframe}
}
\noindent Conversions from flows to FlowGroups as well as mapping of path allocations to pre-established TCP flows between datacenters happen internally. 

The job master can check the status of a submitted coflow using its CoflowId.
{
\begin{colframe}
	\fontsize{9.5pt}{11.4pt}
\begin{Verbatim}
val status = checkStatus(cId)
\end{Verbatim}
\end{colframe}
}

Finally, the job master can modify flows within a submitted coflow as well. 
{\name} assumes that flows within a coflow are uniquely identifiable.
{
\begin{colframe}
	\fontsize{9.5pt}{11.4pt}
\begin{Verbatim}
updateCoflow(cId, Flows)
\end{Verbatim}
\end{colframe}
}

\paragraph{Integration with Apache YARN}
Essentially, {\name} substitutes the Shuffle Service of YARN.
{\name} master runs in the same datacenter as the Application Master of GDA jobs.
When Map tasks finish, instead of a shuffle request, the Application Master submits a coflow request to the {\name} controller.
{\name} informs the Application Master after the coflow has completely been transferred.

\section{Evaluation}
\label{sec:eval}

We evaluated {\name} on $3$ WAN topologies and $4$ benchmarks/industrial workloads in testbed experiments and large-scale simulations. 
Our key findings are as follows:
\begin{denseitemize}
	\item In experiments, {\name} improves the average JCT by $1.55$--$3.43\times$ on average and $2.12\times$--$8.49\times$ at the {\ninefive} percentile w.r.t. TCP while improving WAN utilization by $1.32\times$--$1.76\times$ (\S\ref{sec:eval-exp}).
   
	\item Extensive simulations show that {\name}'s average benefits across 12 <WAN topology, workload> combinations range from $1.04\times$--$2.53\times$ in the smallest topology and $1.52\times$--$26.97\times$ in the largest topology. 
  {\ninefive} percentile improvements are similar (\S\ref{sec:eval-sim}). 
   
   \item {\name} enables $2.82\times$--$4.29\times$ more coflows to meet their deadlines in our testbed experiments (\S\ref{sec:eval-deadline}).
   
   \item {\name} is robust against failures (\S\ref{sec:eval-failure}), its controller can scale to large topologies (\S\ref{sec:eval-scalability}), and it performs well under different parameter settings (\S\ref{sec:eval-sensitivity}).

\end{denseitemize}
\subsection{Methodology}
\paragraph{WAN Topologies}
We consider $3$ inter-DC WANs.

\textit{1. SWAN} \cite[Figure~8]{swan}: Microsoft's inter-datacenter WAN with $5$ datacenters and $7$ inter-datacenter links. 
We calculate link bandwidth using the setup described by Hong {\etal} \cite{swan}. %

\textit{2. {\B4}} \cite[Figure~1]{b4}: Google's inter-datacenter WAN with $12$ datacenters and $19$ inter-datacenter links. 

\textit{3. ATT} \cite{att-mpls}: AT\&T's MPLS backbone network in North America with $25$ nodes and $56$ links. 
We consider one datacenter connected to each node to create a topology larger than SWAN and {\B4}.

Given the locations, we use geographic distances as proxies for link latencies. 
Similar to Hong {\etal} \cite{swan}, we estimate capacities for {\B4} and ATT using the gravity model \cite{gravity-model}. 

\paragraph{Workloads}
We consider $4$ workloads that consist of mix of jobs from public benchmarks -- \emph{TPC-DS}~\cite{tpc-ds}, \emph{TPC-H}~\cite{tpc-h}, and \emph{BigBench}~\cite{bbench} -- and from \emph{Facebook (FB)} production traces \cite{coflow-benchmark, swim}. 
In each run for the benchmarks, jobs are randomly chosen from one of the corresponding benchmarks and follow an arrival distribution similar to that in production traces, because the benchmarks do not have arrival distributions. 
We vary the scale factor of the queries from $40$ to $100$, so each job lasts from few minutes to tens of minutes. 
Each benchmark experiment has $400$ jobs, where job DAGs are generated by the Apache Calcite query optimizer \cite{calcite} during execution by Tez \cite{tez}.
The Facebook one has $526$ simple MapReduce jobs. 
Input tables for jobs are placed across datacenters in a way that a single table can spread across at most $\frac{N}{2}+1$ out of $N$ datacenters. 
Tasks run with datacenter locality.

\paragraph{Testbed Setup}
We built a testbed to emulate the SWAN topology \cite{swan} with $10$ machines in each datacenter. 
{\name} controller runs on a host inside the datacenter that minimizes the control message latency to all datacenters.
Each datacenter has one switch, represented by a machine running one Open vSwitch instance.
The switches are connected by VLANs, on top of a physical 40Gbps switch.
We set the capacity of the links between switches to 1Gbps, to avoid saturating the physical switch.
The bandwidth and latency constraints are enforced by Linux Traffic Control (\texttt{tc}).

\paragraph{Simulator}
We conducted large-scale evaluations using a flow-level simulator that has the same logic as in the actual {\name} controller.
The simulator assumes instant communication between {\name} components. 

\paragraph{Baselines}
We compare {\name} against five baselines:
\begin{denseenum}
	\item \emph{Per-Flow Fairness}: An ideal, single-path per-flow fairness scheduler.
    Flows follow fixed routes calculated by the controller.
    We use TCP in experiments.
	\item \emph{Multipath}: An ideal multipath extension to Per-Flow. 
	\item \emph{SWAN-MCF}: WAN optimizer proposed in \cite{swan}.
	\item \emph{Varys}: Coflow scheduler proposed in \cite{varys}. 
  \item \emph{Rapier}: Coflow scheduling-routing solution for datacenters proposed in\cite{rapier}. 
    We choose $\delta=20$, because it performed the best among values from $\delta=5$ to $\delta=100$.
\end{denseenum}

\paragraph{Metrics}
Our primary metric to quantify performance is the improvement in the \emph{average JCT} computed as:
\begin{equation*}
\text{Factor of Improvement} = \frac{\text{Duration of an Approach}}{\text{Duration of {\name}}}
\end{equation*}
Factor of Improvement (FoI) greater than $1$ means {\name} is performing better, and vice versa.

We use FoI in \emph{average WAN utilization} across the entire WAN (calculated using the same method as above) to compare the efficiency of the compared solutions.

For deadline-sensitive coflows, the primary metric is the percentage of coflows that meet their deadlines.

We use $k=15$ and $\alpha=0.1$ as defaults.

\subsection{{\name}'s Performance in Testbed Experiments}
\label{sec:eval-exp}
We evaluated {\name} on our testbed for SWAN topology to examine its impact on JCT and WAN utilization.
We evaluate {\B4} and ATT topologies using simulation in Section~\ref{sec:eval-sim}.

\begin{figure}[!t]
	\centering
	\subfloat[][JCT]{%
		\label{fig:exp-foi-jct}%
		\includegraphics[width=0.5\linewidth]{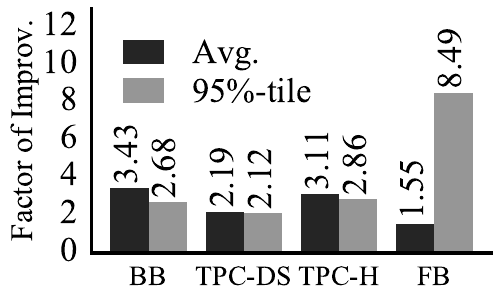}%
	}
	\hfill
	\subfloat[][CCT]{%
		\label{fig:exp-foi-cct}%
		\includegraphics[width=0.5\linewidth]{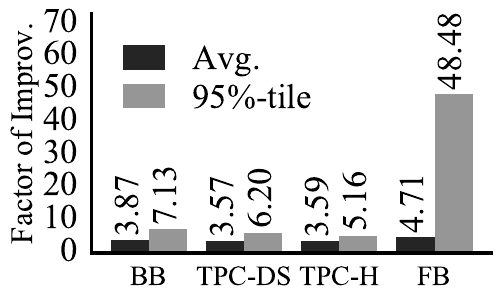}%
	}
\vspace{-3mm}
	\caption{[Testbed] Factors of Improvement of JCT and CCT using {\name} w.r.t. Per-Flow for different workloads on SWAN.}
	\label{fig:exp-foi}
\end{figure}

\begin{figure*}[!t]
\vspace{-0.2cm}
	\centering
	\subfloat[][BigBench]{%
		\label{fig:exp-jct-swan-bb}%
		\includegraphics[width=0.25\linewidth]{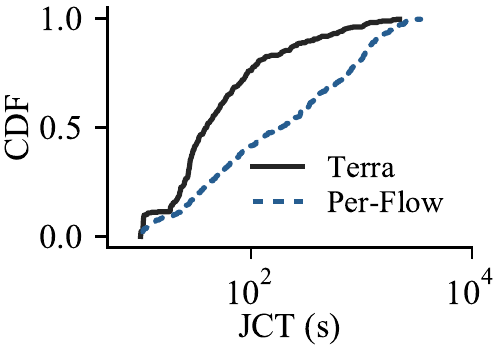}%
	}
	\hfill
	\subfloat[][TPC-DS]{%
		\label{fig:exp-jct-swan-tpcds}%
		\includegraphics[width=0.25\linewidth]{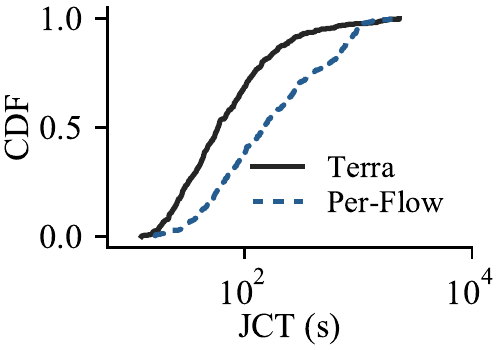}%
	}
	\hfill
	\subfloat[][TPC-H]{%
		\label{fig:exp-jct-swan-tpch}%
		\includegraphics[width=0.25\linewidth]{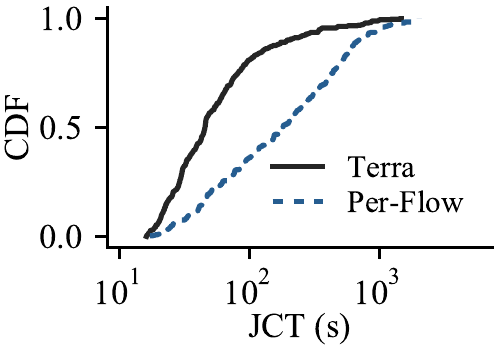}%
	}
	\hfill
	\subfloat[][FaceBook]{%
		\label{fig:exp-jct-swan-fb}%
		\includegraphics[width=0.25\linewidth]{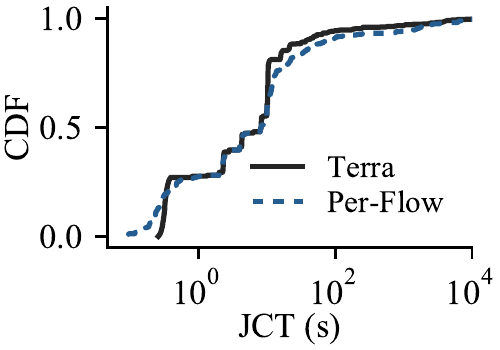}%
	}
  \vspace{-0.2cm}
	\caption{[Testbed] JCTs of individual jobs using {\name} and Per-Flow for different workloads on the SWAN topology.}
	\label{fig:exp-jct}
  \vspace{-0.3cm}
\end{figure*}

\subsubsection{Impact on JCT}
Figure~\ref{fig:exp-foi-jct} shows that {\name} improves the average JCT by at least $1.55\times$ in comparison to Per-Flow (\ie, single-path, fixed-route TCP). 
At the {\ninefive} percentile, the factor of improvement is at least $2.12\times$. 
Note that these numbers include the overheads of schedule computation, preemption, and rescheduling messages from the controller. 
Figure~\ref{fig:exp-foi-cct} shows the improvements in the average CCT, which does not include computation time and are noticeably higher. 
Figure~\ref{fig:exp-jct} presents the CDFs of JCTs for all jobs in all $4$ workloads.

We observe that the FB workload exhibits a different improvement factor than others. 
This is because the FB trace \cite{coflow-benchmark} has heavily skewed distributions -- most jobs have little to no traffic, while a few have most of the tasks and account for almost all the volume. 
This is consistent with prior observations \cite{varys}.
Because of heavy skews in both the number of flows and flow sizes, scheduling and multipath routing enabled by {\name} provide even bigger benefits than that observed by Chowdhury {\etal} \cite{varys} for a longer version of the same trace. 
At the same time, however, Figure~\ref{fig:exp-jct-swan-fb} shows that {\name} does not perform well when scheduling sub-second jobs; we advise not to centrally schedule such small jobs.

\subsubsection{Impact on WAN Utilization}

As is shown in Table~\ref{tab:exp-util-foi}, {\name} improves the WAN utilization by at least $1.32\times$ by effectively using multiple paths.

\begin{table}[!t]
	\begin{center} 
		\begin{small}
			\begin{tabular}{c c c c c }
				\hline \hline
				Workload & BigBench  & TPC-DS   &TPC-H& FB\\
				\hline
				FoI & 1.76   & 1.49   & 1.32 & 1.64\\
				\hline \hline
			\end{tabular}
		\end{small}
	\end{center}
	\caption{[Testbed] WAN utilization FoI of {\name} w.r.t. Per-Flow.}
	\label{tab:exp-util-foi}
\end{table}

\subsection{{\name} in Trace-Driven Simulations}
\label{sec:eval-sim}
So far we have compared {\name} only to single-path, fixed-route TCP. 
A natural question is whether {\name}'s improvements are mostly due to scheduling coflows or from multipath routing. 
Here, we extend our evaluation to simulate and compare {\name} across all $12$ <topology, workload> combinations against five baselines that focus on either scheduling or routing. 
We assume 100 machines per datacenter in these cases.

\begin{table*}[!t]
	\begin{center} 
		\begin{footnotesize}
			\begin{tabular}{c c|r r r r|r r r r|r r r r  }
				\hline \hline
				\multirow{2}{*}{ }	&Topology 		& \multicolumn{4}{c|}{SWAN} & \multicolumn{4}{c|}{{\B4}} & \multicolumn{4}{c}{ATT} \\
				\cline{2-14}	%
				
				&	Workload			& BB   & FB   &TPC-DS& TPC-H& BB   & FB   &TPC-DS& TPC-H& BB   & FB   & TPC-DS & TPC-H \\
				\hline
				\multirow{2}{*}{Per-Flow}&Avg. & 1.78 & 1.96 & 2.16 & 2.53 & 9.79 & 8.08 & 12.45 & 16.63 & 8.64 & 4.50 & 8.98 & 9.57 \\
				&	95\%-tile & 1.57 & 10.04 & 1.63 & 1.65 & 8.82 & 86.29 & 13.40 & 11.93 & 10.61 & 74.66 & 7.41 & 7.15 \\
				\hline
				\multirow{2}{*}{Varys}&Avg. & 1.15 & 1.04 & 1.16 & 1.68 & 6.46 & 3.87 & 8.03 & 16.32 & 10.05 & 2.79 & 11.44 & 26.97 \\
				&	95\%-tile & 1.02 & 1.67 & 1.01 & 1.81 & 5.08 & 2.73 & 13.24 & 25.17 & 11.04 & 5.25 & 12.86 & 30.94 \\
				\hline
				\multirow{2}{*}{SWAN-MCF}&Avg. & 1.59 & 1.69 & 1.73 & 2.17 & 6.13 & 4.26 & 8.81 & 12.60 & 6.64 & 2.64 & 7.11 & 7.83 \\
				&	95\%-tile & 1.28 & 8.80 & 1.27 & 1.34 & 4.98 & 42.85 & 9.40 & 9.04 & 7.18 & 38.49 & 5.34 & 5.29 \\
				\hline
				\multirow{2}{*}{Multipath}&Avg. & 1.30 & 1.65 & 1.25 & 1.28 & 2.77 & 4.11 & 2.88 & 3.24 & 3.75 & 2.48 & 3.03 & 2.82 \\
				&	95\%-tile & 1.15 & 7.66 & \cellcolor{gray!20}0.90 & \cellcolor{gray!20}0.86 & 2.49 & 40.99 & 3.51 & 2.96 & 5.24 & 34.86 & 2.32 & 2.26 \\
				\hline
				\multirow{2}{*}{Rapier}&Avg. & 1.67 & 2.15 & 1.32 & 2.09 & 1.80 & 1.91 & 1.81 & 1.80 & 2.37 & 1.90 & 1.73 & 1.52 \\
				&	95\%-tile & 1.64 & 2.70 & 1.25 & 1.57 & 1.63 & 5.72 & 1.60 & 1.88 & 2.64 & 4.23 & 1.55 & 1.42 \\
				\hline \hline
			\end{tabular}
		\end{footnotesize}
	\end{center}
	\caption{[Simulation] Factors of improvement in the average and {\ninefive} percentile JCT using {\name} w.r.t. baselines.}
	\label{tab:sim-foi}
  \vspace{-0.6cm}
\end{table*}

\subsubsection{Impact on JCT}
Table~\ref{tab:sim-foi} shows that {\name} improves the average JCT by $1.04\times$--$2.53\times$ in SWAN, $1.80\times$--$16.63\times$ in {\B4} and $1.52\times$--$26.97\times$ in ATT with similar {\ninefive} percentile improvements. 
Although we omit detailed CDFs for brevity, {\name} outperforms its counterparts in varying degrees across all percentiles except for the shaded cells in Table~\ref{tab:sim-foi}.

\paragraph{Which Jobs See the Biggest Benefits?}
We calculated the Pearson's correlation coefficients ($r$) between FoIs w.r.t. the baseline and total WAN transfers for all $12$ <topology, workload> combinations. 
The result showed consistent negative correlations ($-0.05$ to $-0.39$) across the board, suggesting that smaller jobs see more benefits than the bigger ones.

\paragraph{How does Topology or Workload Affect?}
There were no significant correlations between {\name}'s improvements and workloads, except for FB that showed a lower improvement for average JCT than that at the {\ninefive} percentile. 
This is again likely due to its heavy-tailed distribution. 
The large number of small jobs receive little benefit due to coordination overheads, resulting in lower improvement on average. 
Speedups from {\name} for larger jobs contribute to the {\ninefive} percentile.

Performances of {\name} and all baselines do vary across topologies. 
{\name} performs increasingly better in larger topologies (\ie, {\B4} and ATT). 
In these cases, Multipath comes the closest to {\name} because it leverages many available paths between datacenters.
For the SWAN topology, Varys is the closest to {\name} because there are not many available paths (only $5$ nodes and $7$ links).
Overall, {\name} outperforms the rest via joint scheduling-routing co-optimization.

\paragraph{How Far are We From the Optimal?}
Calculating the optimal solution of a computationally intractable problem is infeasible. 
Instead, we use slowdowns of coflows as a loose lower bound. 
We calculated slowdown of a coflow by comparing the completion time using {\name} with its minimum (\ie, in an empty network).
We found that the average slowdown using {\name} was between $1.08\times$ and $2.95\times$ across all $12$ <topology, workload> combinations.
In contrast, the baselines ranged between $1.77\times$ and $82.18\times$.

\subsubsection{Impact on WAN Utilization}
Table~\ref{tab:sim-util-foi} shows the improvements in WAN utilization using {\name} w.r.t. the best of all five baselines in terms of utilization for all <topology, workload> combinations. 
{\name} achieved $1.06\times$ to $1.76\times$ higher utilization for all combinations except <SWAN, FB>.

\begin{table}[!t]
	\begin{center} 
		\begin{small}
			\begin{tabular}{c c c c c }
				\hline \hline
				Workload & BigBench  & TPC-DS   &TPC-H& FB\\
				\hline
				SWAN 	& 1.12  & 1.14  & 1.06 & 0.92 \\
				G-Scale & 1.22	& 1.14	& 1.09 & 1.12 \\
				ATT		& 1.42	& 1.34	& 1.76 & 1.38 \\
				\hline \hline			
			\end{tabular}
		\end{small}
	\end{center}
	\caption{[Simulation] WAN utilization FoI of {\name}.}
	\label{tab:sim-util-foi}
\end{table}

\subsection{Performance on Deadline-Sensitive Coflows}
\label{sec:eval-deadline}
For deadline-constrained experiments, we set the deadline of a coflow to be $d\times$ its minimum completion time (\ie, in an empty network) and vary $d$ from $2$ to $6$. 
We run experiments for BigBench and SWAN in both Testbed and simulation.

\begin{figure}[!t]	
	\centering
	\includegraphics[scale=0.9]{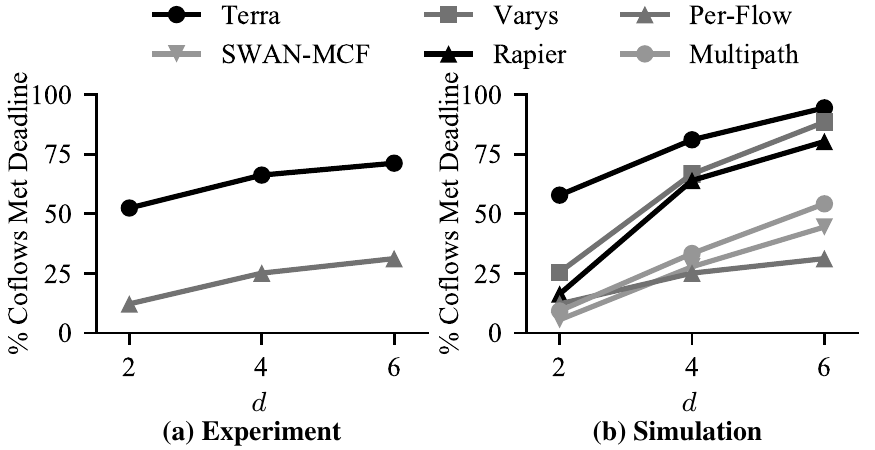}
	\vspace{-3mm}
	\caption{Percentage of coflows that meet deadline w.r.t the baselines for deadlines set to $d\times$ of a coflow's minimum CCT.}
	\label{fig:deadline}
\end{figure}

\paragraph{Testbed Experiments} 
Figure~\ref{fig:deadline}a shows that {\name} allowed $2.82\times$ to $4.29\times$ more coflows to complete within deadlines compared to Per-Flow.
Note that a small fraction of the admitted coflows missed their deadlines. 
This is due to the uncertainty of the global state at the controller in between feedback loops.
However, most of them still completed within $50\%$ of their deadline.

\paragraph{Trace-driven Simulations} 
Figure~\ref{fig:deadline}b shows that {\name} allowed $1.07\times$ to $2.31\times$ more coflows to complete within deadlines compared to all the baselines in simulations.
Here, all admitted coflows completed in {\name} because we assume instantaneous communication between the controller and agents.

\subsection{Reactive Re-Optimization Upon Failure}
\label{sec:eval-failure}

\begin{figure}[!t]
		\vspace{-7mm}
	\centering
	\subfloat[][]{%
		\label{fig:exp-failure-a}%
		\includegraphics[width=0.25\linewidth]{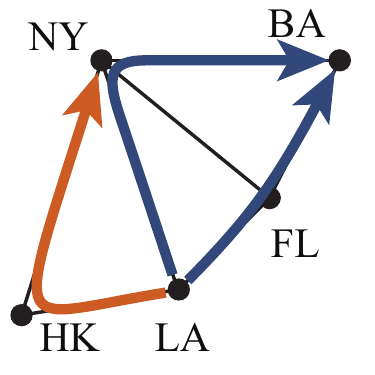}%
	}
	\hfill
	\subfloat[][]{%
		\label{fig:exp-failure-b}%
		\includegraphics[width=0.25\linewidth]{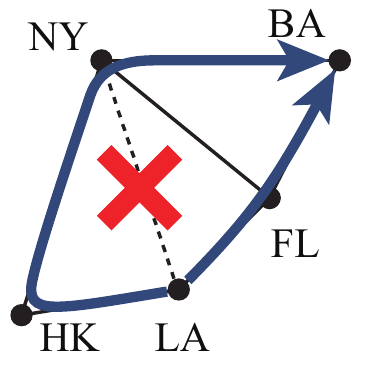}%
	}
	\hfill
	\subfloat[][]{%
		\label{fig:exp-failure-c}%
		\includegraphics[width=0.25\linewidth]{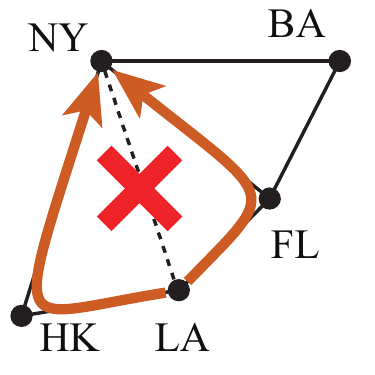}%
	}
	\hfill
	\subfloat[][]{%
		\label{fig:exp-failure-d}%
		\includegraphics[width=0.25\linewidth]{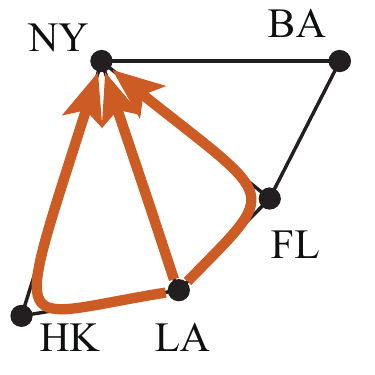}%
	}
\vspace{-3mm}
	\caption{[Testbed] An example of failure handling. FlowGroup of Job~$1$ is denoted by the blue (dark) arrows, FlowGroup of Job~$2$ is denoted by orange (light) arrows. The dashed line with a cross denotes the failed link. (b): Link failed, Job~$2$ got preempted. (c): Job~$1$ finished, Job~$2$ got rescheduled. (d): Link recovered, Job~$2$ received a new path.}
	\label{fig:exp-failure-example}
	
\end{figure}

\begin{figure}[!t]
	\centering
	\includegraphics[scale=0.8]{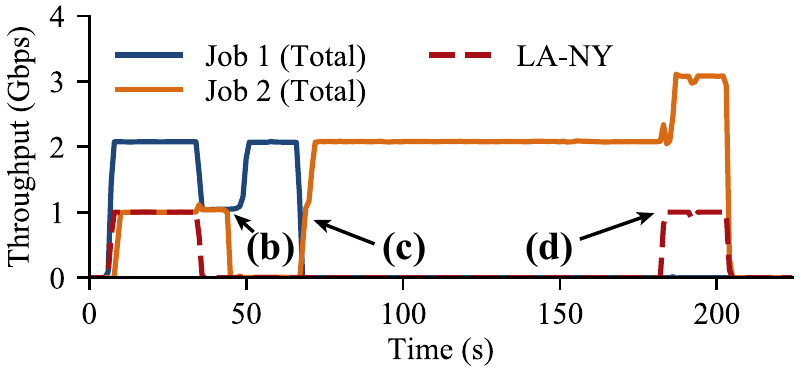}%
	\caption{[Testbed] Throughput of jobs and the failed link.}
	\label{fig:exp-failure-util}
\end{figure}

We evaluate {\name}'s robustness upon failure by a case study in our testbed. 
Figure~\ref{fig:exp-failure-example} shows an example where one link (LA-NY) failed when there are two jobs \{Job~$1$, Job~$2$\} running, and Figure~\ref{fig:exp-failure-util} shows the throughput change throughout this process.
Note that we set $\alpha=0$ for ease of exposition. 

In this case, Job~$1$ has a smaller volume and thus higher priority than Job~$2$.
{\name} reacts to the link failure within 10 seconds and preempts Job~$2$ (Figure~\ref{fig:exp-failure-b}), minimizing the impact of link failure to Job~$1$'s throughput.
After the completion of Job~$1$, {\name} re-schedules Job~$2$ (Figure~\ref{fig:exp-failure-c}). 
Finally, when the failed link is reinstated (Figure~\ref{fig:exp-failure-d}), {\name} adds a new path for Job~$2$.

\subsection{{\name} Overheads}
\label{sec:eval-scalability}
\paragraph{Computational Complexity}
In our experiments, we ran {\name} controller on a machine with two 2.6GHz Intel Xeon Gold 6142 CPU and measured the number of LP computations and total time spent for each scheduling decision.
For BigBench on SWAN topology, for each schedule, {\name} needs to solve $28.4$ LPs on average, taking $74.43$~ms. 
For the same workload on ATT topology, each schedules takes $589.1$~ms to solve $31.46$ LPs on average ($2.991$~s to solve $52$ LPs at the {\ninefive} percentile).
Given that many job lasts for a few minutes, this overhead is acceptable in most cases. 

\begin{figure}[!t]
\vspace{-0.1cm}
	\centering
	\includegraphics[width=0.6\linewidth]{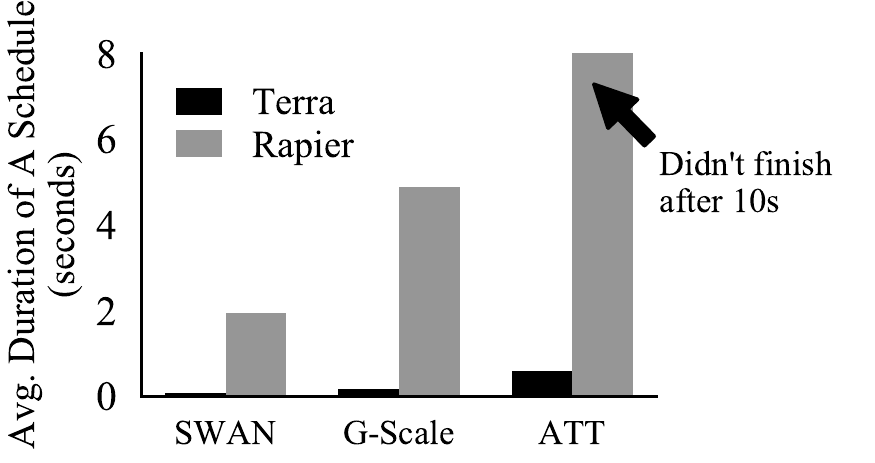}%
	\vspace{-0.3cm}
	\caption{Scheduling overhead of {\name} against Rapier. }
	\label{fig:overhead-comparison}
\end{figure}

We also compared {\name} against Rapier, as shown in Figure~\ref{fig:overhead-comparison}. 
For Rapier optimizing BigBench on SWAN topology, each scheduling round takes $1.952$ seconds to solve $36.5$ LPs, which is $26.2\times$ worse than {\name}. 
For a larger topology (G-Scale), Rapier is even worse ($29.14\times$).
This is because FlowGroups reduce the number of flows by up to $100\times$ when geo-distributed tasks are spread across $10$ hosts in each datacenter.

\paragraph{Number of Rules and Rule Updates}
{\name} incurs rule setup cost upon starting and in case of failures. 
{\name} installs only up to $168$ flow rules in each switch for the SWAN topology.
Operators can vary $k$ to limit the number of rules per switch.

\subsection{Sensitivity Analysis}
\label{sec:eval-sensitivity}

\begin{figure}[!t]
	\centering
	\subfloat[][Improvements in Avg. JCT]{%
		\label{fig:sim-ksp-jc}%
		\includegraphics[width=0.5\linewidth]{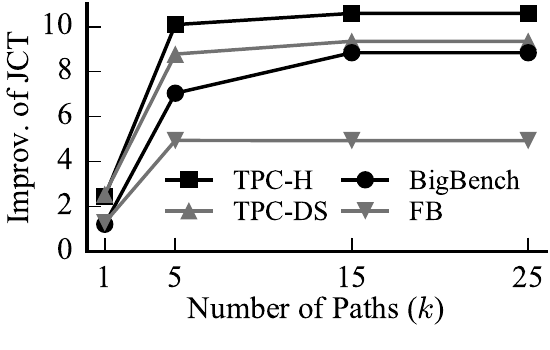}%
	}
  \hfill
	\subfloat[][Improvements in Utilization]{%
		\label{fig:sim-ksp-bw}%
		\includegraphics[width=0.5\linewidth]{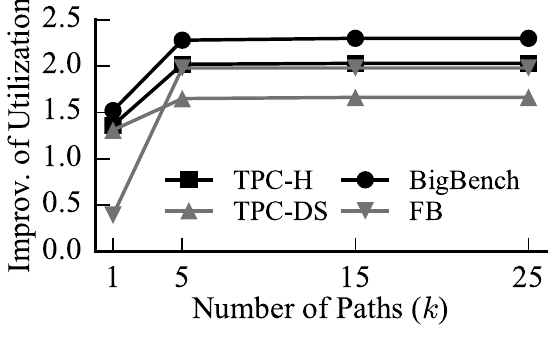}%
	}
	\caption{[Simulation] Improvements in the average JCT and utilization using {\name} w.r.t. Per-Flow on ATT topology as the number of paths between two datacenters ($k$) is varied.}
	\label{fig:sim-ksp}
\end{figure}

\paragraph{Impact of Restricting Paths}
{\name} minimizes the number of rules it has to install by reusing existing TCP connections across multiple end-to-end paths between two datacenters. 
However, if the number of possible paths between them ($k$) increases, {\name} has to establish and maintain more TCP connections. 
Consequently, {\name} allows operators to restrict the number of possible paths between datacenters (\S\ref{sec:one-job}). 
Figure~\ref{fig:sim-ksp} shows the impact of $k$ for the all workloads on the ATT topology. 
We observe that: 
(i) With larger $k$, {\name}'s improvements become higher. This is because larger $k$ allows for more multipath benefits. 
(ii) After $k$ reached a threshold (it is between $k=5$ and $k=10$ for ATT), increasing $k$ does not significantly affect JCT or utilization.

\begin{figure}[!t]
	\centering
	\includegraphics[width = 0.7\linewidth]{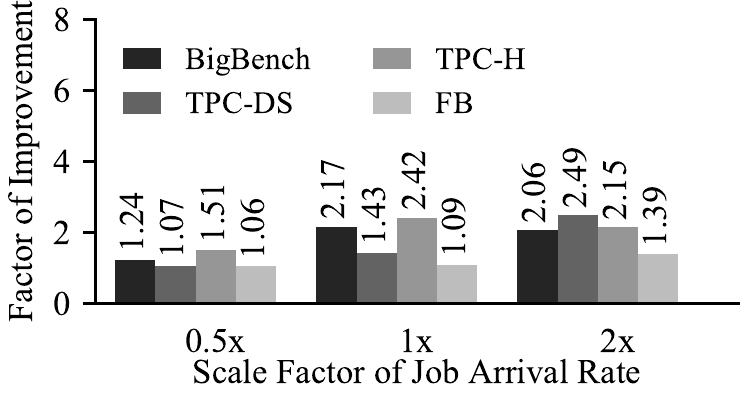}%
		\vspace{-3mm}
	\caption{[Testbed] Factor of Improvement in the average JCT for workloads with scaled arrival rates in the SWAN topology.}

	\label{fig:exp-varyA}
\end{figure}

\paragraph{Impact of Job Arrival Rate and Load}
We scaled the arrival rate of the queries to increase load and evaluated {\name}'s performance in testbed.
Figure~\ref{fig:exp-varyA} shows that {\name} performs better with increasing arrival rate -- \ie, with increasing load. 
When the network is lightly loaded, the room for performance improvement is lower.
Increasing load by making jobs larger (instead of making jobs arrive in shorter intervals) also resulted in increasing benefits.

\begin{figure}[!t]
	\centering
	\includegraphics[scale=0.8]{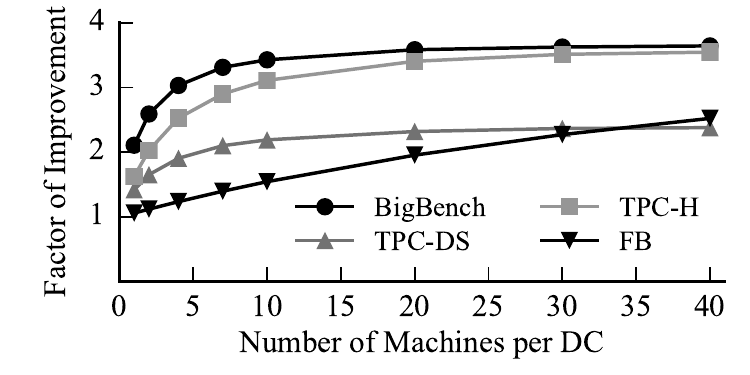}%
  \vspace{-0.3cm}
	\caption{[Testbed] Improvements in the average JCT using {\name} w.r.t. Per-Flow on the SWAN topology with increasing number of machines in each datacenter.}
	\label{fig:sim-varyN}
\end{figure}

\paragraph{Computation v.s. Communication}
By keeping the time spent in communication constant, we vary the number of machines in each datacenter and estimate the average JCT. %
Figure~\ref{fig:sim-varyN} shows improvements of average JCTs w.r.t. the number of machines used for all jobs on SWAN topology.
Because $\mbox{JCT} = (T_{\mbox{Comm}} + T_{\mbox{Comp})}$ for each stage, the improvements increase with the number of machines used.

\paragraph{Choice of $\alpha$}
We also compared $\alpha=0.2$ and $\alpha=0.1$ for BigBench on the SWAN topology and found that the average JCT is $2.3\%$ higher for $\alpha=0.2$.
\section{Related Work}
\label{sec:related}

\paragraph{Geo-Distributed Analytics}
Existing geo-distributed analytics solutions focus on two goals: minimizing WAN usage and minimizing the average JCT via query planning and/or data placement \cite{iridium, geode, jetstream, gaia-cmu, clarinet, swag, lube, tetrium2018}.
However, they all ignore the WAN topology and are not robust to WAN events. 
In contrast, {\name} focuses on simultaneously scheduling and routing WAN coflows and can complement existing solutions. 

\paragraph{Speeding up Data Analytics}
Efforts in speeding up data analytics include scheduling \cite{drf, tetris, carbyne, late, mantri, quincy, mesos}, caching \cite{pacman, ec-cache, spark, tachyon}, query planning \cite{sparksql, shark, calcite}, straggler mitigation \cite{dolly, mantri, kmn, delay-scheduling}, approximation \cite{blinkdb, grass}, and data placement \cite{sinbad, cohadoop, gfs, fds}.
The primary context along these directions is a single datacenter, whereas {\name} focuses on optimizing performance across multiple datacenters, where WAN bandwidth is often the primary bottleneck.

\paragraph{Flow and Coflow Scheduling}
Flow scheduling \cite{pdq, hedera, dctcp, detail, pfabric, pias} and coflow scheduling \cite{orchestra, varys, aalo, coda, sincronia} deal with optimizing communication performance in datacenters. 
The former is application-agnostic, while the latter is application-aware but assumes full bisection bandwidth datacenter networks. 
Similar to them, {\name} can account for minimizing completion times as well as meeting deadlines; unlike them, {\name} is both topology- and application-aware. 

Rapier \cite{rapier} comes the closest to {\name} in that it also considers routing alongside scheduling to handle datacenter topologies.
In contrast, {\name} considers general WAN topologies. 
Furthermore, Rapier uses single-path routing, does not use {\name}'s FlowGroup optimization, relies on time-division multiplexing (using its $\delta$ parameter) to avoid starvation, and updates switch rules on every reschedule -- all of which contribute to its lower performance and higher complexity (\S\ref{sec:eval-sim}).

Siphon \cite{siphon} is a WAN-agnostic framework minimizing CCT for GDA jobs.
Although it applies similar ideas of creating a overlay network and multipath transfers on that overlay, it is agnostic to the underlying WAN.
As such, it does not consider WAN routing and cannot directly react to WAN uncertainties.

\paragraph{WAN Traffic Engineering}
Traditionally, optimizing WAN transfers revolved around tuning (ECMP and/or OSPF) weights \cite{te-ip, te-ospf} and adapting allocations across pre-established tunnels \cite{tightrope, cope}, often via MPLS \cite{mpls-te}.
With the advent of SDNs, Google \cite{b4, bwe} and Microsoft \cite{swan} have shown that it is indeed possible to perform traffic engineering in a (logically) centralized manner. 
{\name} builds on top and extends the latter works. 
Another relevant body of work is WAN data transfers with deadlines.
However, unlike these point-to-point \cite{bwe, tempus, amoeba, pretium, netstitcher, owan} or point-to-multipoint \cite{dccast} solutions, {\name} leverages application-level semantics to optimize multipoint-to-multipoint coflows.
\section{Conclusion}
Despite growing interests in geo-distributed analytics and SD-WANs, there exists a large gap between the two emerging areas. 
The former ignores the WAN, while the latter ignores readily available application-level semantics, leading to large performance loss.
{\name} bridges this gap by enabling scheduling-routing co-optimization in geo-distributed analytics via SDN-based WAN traffic engineering. 
It uses coflows to schedule WAN transfers from geo-distributed jobs, and leverages FlowGroups and SDN to compute and dictate their routes across multiple paths. 
Integrations with the FloodLight SDN controller and Apache YARN, and evaluation on $12$ <topology, workload> combinations show that co-optimization can improve the average JCT by $1.55\times$--$3.43\times$ on average while moderately improving the average WAN utilization and complete $2.82\times$--$4.29\times$ more coflows within their deadlines.
Furthermore, {\name} enables geo-distributed jobs to dynamically react to large bandwidth fluctuations due to failures in the WAN and traffic variabilities.

 In conclusion, this paper is only a first step in bringing WAN into the picture of geo-distributed analytics.
 It opens up several exciting research problems, which include co-optimizing without knowing the WAN topology (\eg, when running in the cloud environment), without knowing the traffic matrices (\eg, geo-distributed streaming), and extending to streaming analytics.
\phantomsection
\label{EndOfPaper}
\end{sloppypar}

{
\bibliographystyle{ACM-Reference-Format}
\bibliography{gaia}
}

\end{document}